\documentclass[11pt,a4paper]{JHEP3_mod2}

\usepackage{amsmath,latexsym,amssymb,slashed}
\usepackage{graphicx}				
\usepackage{amssymb}					
\usepackage{tabularx}				
\usepackage[vcentermath]{youngtab}	

\usepackage[latin1]{inputenc}
\usepackage[T1]{fontenc}

\newcommand{\eq}[1]{\begin{equation}#1\end{equation}}
\newcommand{\spl}[1]{\begin{split}#1\end{split}}

\newcommand{\D}{\mathcal{D}}
\newcommand{\p}{\mathcal{P}}

 \DeclareMathOperator{\rank}{rank}

\def\d{\text{d}}

\newcommand{\cp}[1]{\mathbb{CP}^{#1}}

\author{Robin Terrisse and Dimitrios Tsimpis\\
Universit\'{e} Claude Bernard (Lyon 1)\\
Institut de Physique Nucl\'{e}aire de Lyon, UMR 5822, CNRS/IN2P3\\
4 rue Enrico Fermi,
F-69622 Villeurbanne Cedex,  France\\

E-mail:
\email{terrisse@ipnl.in2p3.fr}, \email{tsimpis@ipnl.in2p3.fr}}
\abstract{We construct 
globally-defined $SU(3)$  structures on smooth compact toric varieties (SCTV) in the class of $\mathbb{CP}^1$ bundles over $M$, where $M$ is an arbitrary  
 SCTV of complex dimension two. The construction  can be extended to the case where the base is K\"{a}hler-Einstein of positive curvature, but not necessarily toric, and admits a parameter space which includes $SU(3)$  structures of LT type.}
\title{SU(3) structures on S\boldmath$^2$\unboldmath~{}bundles over four-manifolds
 }
\preprint{}

\begin{document}
\setlength{\parindent}{0pt}



\section{Introduction}

In many cases the problem of supersymmetric compactification to four-dimensional Minkowski or AdS space can be reformulated as the problem of existence of $SU(3)$ structures with appropriate torsion classes parameterized by the ``fluxes''. Although this approach has lead to great progress in the construction of string vacua, 
 the search for manifolds with suitable $SU(3)$ structures has been far less systematic  than the construction of Calabi-Yau manifolds, for which powerful algebro-geometric tools are 
 available. 
 
 In \cite{Larfors:2010wb} it was proposed to use smooth compact toric varieties (SCTV) as a class of manifolds 
 for which tools from both algebraic and differential geometry can be used, and  develop a formalism 
suitable for the description of $SU(3)$ structures on SCTV. The idea is to use the canonical structure that comes with the symplectic quotient description of 
 the SCTV (metric, complex structure, set of coordinates), and construct on it a different (nonintegrable in general) almost complex structure associated with a globally-defined $SU(3)$ structure.

The construction of  $SU(3)$  structures on SCTV proposed in \cite{Larfors:2010wb} relies on the existence of a one-form $K$ in the 
parent space of the symplectic quotient, satisfying certain requirements. Thus the problem of constructing  $SU(3)$  structures on SCTV is reduced 
to the problem of constructing one-forms $K$ satisfying the requirements of \cite{Larfors:2010wb}.  Although that reference gave some examples 
of suitable one-forms, and many more were subsequently constructed in \cite{Larfors:2013zva}, no general formula for $K$ exists 
satisfying the requirements of \cite{Larfors:2010wb}.  As a result, the search for $SU(3)$  structures on SCTV had up to now proceeded on a case-by-case basis.

In the present paper 
we extend the formalism of \cite{Larfors:2010wb} for SCTV to construct 
globally-defined $SU(3)$  structures on  the class $\mathbb{CP}^1$ over $M$, where $M$ is an arbitrary  two-dimensional SCTV. 
As in \cite{Larfors:2010wb}, our construction is based on the existence of a one-form $K$ which,  in our case, is naturally distinguished 
by the structure of the bundle. This one-form does not have the right $U(1)$ charge (in symplectic-quotient terminology) for the 
procedure of \cite{Larfors:2010wb} to go through. A different procedure is used instead, exploiting the local $SU(2)$ structure of the 
base $M$ of the fibration.

More specifically we give a general formula, eq.~(\ref{main}) below, for globally-defined $SU(3)$ structures on all $\mathbb{CP}^1$ bundles whose  $U(1)$ charges 
satisfy eq.~(\ref{chc}). The latter equation can always be satisfied for any two-dimensional SCTV base, and amounts to choosing a specific twisting of the 
$\cp1$ bundle. The $SU(3)$ structures thus constructed admit a space of deformations parameterized by certain functions, described below eq.~(\ref{233mod}). 
The associated torsion classes depend on these functions, and are nonvanishing in general.

This method can also be used to construct $SU(3)$ structures on $S^2$ bundles over $B_4$, where $B_4$ is  K\"{a}hler-Einstein, but not necessarily toric.  
Provided $B_4$ has positive scalar curvature, {\it i.e.}~if it is $\cp1\times\cp1$, $\cp2$, or one of the del Pezzo surfaces $dP_3,\dots,dP_8$ \cite{Tian:1987if}, the total space of the $S^2$ bundle is  complete and the associated metric is regular. Moreover the parameter space 
includes $SU(3)$  structures of LT type, suitable for supersymmetric AdS$_4$ compactifications of massive IIA.

The outline of the remainder of paper is as follows. In section \ref{sec:formalism} we review the formalism of \cite{Larfors:2010wb} for SCTV, and introduce the tools 
that will be used in the rest of the paper. The toric $\cp1$ 
bundles are described in section \ref{sec:toricb}. In section \ref{sec:3} we work out in detail the example of $\cp1$ over $\cp2$. This is the simplest example in the 
class of toric $\cp1$ bundles over $M$, but it already captures the main idea of the construction. 
The $SU(3)$ structure is constructed in section \ref{sec:spc}. Section \ref{sec:general} discusses the construction of $SU(3)$ structures on toric $\cp1$ bundles over 
general two-dimensional SCTV. Section \ref{sec:lt} discusses the construction of $SU(3)$ structures on $S^2$ bundles over four-dimensional K\"{a}hler-Einstein bases. 
We conclude in section \ref{concl}. 
For ease of presentation, many technical details have been moved to the appendices.


\section{Review of the formalism}\label{sec:formalism}

In order to fix the notation 
and make the paper self-contained, 
in this section we give a review of the SCTV formalism  developed in \cite{Larfors:2010wb}. Along the way we introduce the tools that will be useful in the rest of the paper. 
The description of the toric $\cp1$ bundles is given at the end of the section.

There are various equivalent ways to define a toric variety see {\it e.g.}~\cite{fulton}, or \cite{Denef:2008wq} for an introduction for physicists. 
In the following we will use the symplectic quotient description, which turns out to be the best suited 
for the explicit construction of $G$-structures and the associated differential calculus. The starting point of the symplectic quotient description is a {\it parent space}  
$\mathbb{C}^k$, with coordinates $\{z_i,~i=1,\dots, k\}$, and a set of  $s$  linearly-independent integer $k$-vectors $Q^a_i$, $\{a=1,\cdots,s\}$ called the {\it charges}. 
Let $\tilde{M}$ be the real submanifold defined by the following set of {\it moment map equations},
\begin{equation}\label{struc}
Q_i^a |z_i|^2 = \xi^a~.
\end{equation}
The real parameters $\xi^a$ are the so-called, \textit{Fayet-Iliopoulos} parameters: they correspond to K\"{a}hler moduli, parametrizing the sizes 
of cycles of the toric variety. On the other hand the topology of the variety is independent of the $\xi^a$ as long as we stay inside the {\it K\"{a}hler cone}, 
defined by the conditions $\xi^a>0$. In the following we will always assume this to be the case. The associated  toric variety  $M$ is given by the
 quotient $M=\tilde{M}/U(1)^s$ where the {\it phase vector}   $\phi_a\in U(1)^s$ acts on the coordinates $z_i\in \tilde{M}$ through the following 
 {\it gauge transformations},
\begin{equation}\label{u1}
z_i\rightarrow\phi\cdot z_i := e^{i Q^a_i \phi_a}z_i~.
\end{equation}
Hence $M$ is a manifold of complex dimension $d=k-s$: the equations (\ref{struc}) can be thought of as removing $s$ real ``radial'' directions, whereas the 
action of (\ref{u1}) removes $s$ real `angular' directions. In total the equations (\ref{struc}), (\ref{u1}) remove $s$ pairs consisting of one radial and one angular variable, 
which may be thought of as  $s$ complex variables. 

Since the  $Q^a$ are independent as $k$-vectors, one may choose a set  $S$ of $s$ indices such that  $Q^a_b$, $b\in S$,  is invertible. 
The open set $\left\{z_b\neq 0, b\in S\right\}\subset\mathbb{C}^k$ 
then descends to a well-defined open set in  $M$, denoted by $U_S$. 
On this patch one can then use the  $z_b$ coordinates to compensate the $U(1)^s$ action on the  $z_\alpha$ coordinates, where the index $\alpha$  
takes values in the complement of $S$, $\alpha\in {}^\complement{}S$. 
One may then define the following gauge-invariant quantities,
\begin{equation}\label{local}
t_i:=  z_i \prod_{a\in S} z_a^{-\mathcal{Q}_b^a Q_i^b} ~,
\end{equation}
where we have set,
\eq{\label{covfefe}
\mathcal{Q}^a_b := (Q^a_b)^{-1}~.
}

Thus, provided $|Q_S|:=|\det Q_{b\in S}^a| = 1$, the map,
\begin{equation}\label{chart}
\begin{array}{rcl}
\varphi_S : U_S &\rightarrow & \mathbb{C}^d \\
        \left[ z_i\right] & \mapsto    & (t_\alpha)_{\alpha\in {}^\complement{}S}~,
\end{array}
\end{equation}
where $d:=k-s$, 
is a well-defined homeomorphism, while the transition functions $\varphi_S \circ \varphi_{S'}^{-1}$ are biholomorphic and rational. The charts $(U_S,\varphi_S)$ form a holomorphic atlas on $M=\bigcup_{|Q_S|=1} U_S$: the $t_\alpha$, $\alpha\in {}^\complement{}S$, define $d$ gauge-invariant local holomorphic coordinates on $U_S$. Note that for $i=c\in S$, we thus find $t_c=1$. The existence of this covering of $M$ is related to the condition of smoothness for general toric varieties, and can be checked explicitly for the cases we consider here.

To take a simple example, consider the case  $s=1$ and  $Q=(1,\cdots,1)$. 
The corresponding toric variety is the complex projective space  $\cp{k-1}$. Indeed (\ref{struc}) gives  $\|z\|^2=\xi$, \textit{i.e.} $\tilde{M}= S^{2k-1}$. Taking the  $U(1)$ quotient, 
$M$ can be written as $M=(\mathbb{C}^k\backslash \{0\})/\mathbb{C}^*$,  the set of complex lines in $\mathbb{C}^k$. On the patch $U_j=\left\{z_j\neq 0\right\}$, the local coordinates 
take the form,
\begin{equation}
t_i= \frac{z_i}{z_j}~,
\end{equation}
which we recognize as the set of canonical coordinates of $\cp{k-1}$. The $z_i$ on the other hand correspond to homogeneous coordinates of $\cp{k-1}$.

\subsection{Differential forms}\label{sec:diff}

We have seen that toric varieties are equipped with systems of complex coordinates which can  easily be made explicit. Moreover it is often advantageous to 
work directly in the parent space $\mathbb{C}^k$ using the homogeneous coordinates $z_i$. We will be interested in 
particular in globally-defined differential forms on the manifold $M$. 
One way to construct a differential form on  $M$ is to start from its local expression on a patch, and make sure a regular global extension exists by checking its compatibility with the transition functions of the cotangent bundle. Working directly in  $\mathbb{C}^k$ drastically simplifies this problem: since the topology of the parent space is trivial, a single 
expression suffices to define differential forms globally. 
From this point of view the key question is to identify the differential forms of $\mathbb{C}^k$ which descend to well-defined forms on $M$.

In the following we review how the formalism of \cite{Larfors:2010wb} can be used to treat this question. 
Let $\Phi$ be a differential form on $\mathbb{C}^k$. In order for $\Phi$ to descend to a well defined form on  $M$, it should be well-defined on $\tilde{M}$. Hence it should be compatible with the 
moment map equations (\ref{struc}) which imply, 
\begin{equation}
Q_i^a \bar{z}_i \d{} z_i + Q_i^a z_i \d\bar{z}_i = 0~.
\end{equation} 
Consequently $\Phi$ should not have any components along the $\Re \eta^a$, where we have defined, 
\begin{equation}
\eta^a := Q_i^a \bar{z}_i \d{} z_i~.
\end{equation}
In other words, we require, 
\begin{equation}\label{vertR}
\iota_{\Re (V^a)}\Phi=0,
\end{equation}
where $V^a$ is the dual of $\eta^a$ (with respect to the canonical metric of $\mathbb{C}^k$),
\begin{equation}
V^a := Q_i^a z_i \partial_{z_i}~.
\end{equation}
Moreover, 
$\Phi$ should be compatible with the quotient (\ref{u1}). On the other hand the $U(1)^s$ action in (\ref{u1}) is generated by the  vector fields  $\Im (V^a)$. 
Hence the $U(1)^s$ invariance can be stated in terms of the following two conditions: 
\begin{enumerate}
\item $\Phi$ must be constant along $U(1)^s$ orbits, {\it i.e.} $\mathcal{L}_{\Im(V^a)}\Phi=0$~.
\item $\Phi$ should not have any components along the orbits, {\it i.e.}  $\iota_{\Im (V^a)}\Phi=0$~.
\end{enumerate}
These conditions have a natural interpretation: first note that 
a form  $\Phi$ has charge $q^a$ if it is an eigenvector of the  Lie derivative $\mathcal{L}_{\Im(V^a)}$, 
\begin{equation}
\mathcal{L}_{\Im(V^a)}\Phi= q^a\Phi~.
\end{equation}
We then see that the first of the two conditions above is simply the gauge invariance of $\Phi$, 
 \textit{i.e.} the condition that the total charge of $\Phi$ vanishes. 
Moreover the second condition combined with (\ref{vertR}) gives,
\begin{equation}\label{vert}
\iota_{V^a} \Phi= \iota_{\bar{V}^a}\Phi =0~,
\end{equation}
which is equivalent to $\Phi$ being {\it vertical} with respect to $V^a$. 

Thus in order to construct a well-defined form on $M$ descending from a form $\Phi$ on $\mathbb{C}^k$,  
the gauge invariance of $\Phi$ must be imposed from the outset. On the other hand, the verticality condition 
is purely algebraic and can be imposed by projecting out the components along $\eta^a$. 

Let us now come to 
the explicit construction of the vertical projector. 
We introduce the real symmetric matrix, 
\begin{equation}\label{defg}
g^{ab} := \eta^a(V^b)= Q_i^aQ_i^b |z_i|^2~.
\end{equation}
The projection $P$ of a $(1,0)$-form $\Phi$  is then given by,
\begin{equation}
P(\Phi)=\Phi -\tilde{g}_{ab} \iota_{V^a}(\Phi)\eta^b~,
\end{equation}
where $\tilde{g}=g^{-1}$. This definition of $P$ can be readily extended to  all $(k,l)$-forms \cite{Larfors:2010wb}. 
In the following it will be useful to define the vertical projections, $\mathcal{D}z_i$,  of the one-forms 
$\d z_i$,
\begin{equation}\label{Pdz}
\mathcal{D}z_i:= P (\d{} z_i) = \d{} z_i- \tilde{g}_{ab} Q_j^a Q_i^b \bar{z}_j z_i \d{} z_j = \d{} z_i - h_{ij} z_i \bar{z}_j \d{} z_j~,
\end{equation}
(no sum over $i$)  
where we have set,
\eq{\label{hdef}
h_{ij}:=Q^a_iQ^b_j\tilde{g}_{ab}
~.}
The  $\mathcal{D}z_i$ are the building blocks that we will use to construct global forms on $M$. 
Note however that since they are not gauge invariant, one must compensate their charge by appropriate (charged) coefficients.

On the other hand the (singular) form  ${\mathcal{D}z_i}/{z_i}$ is both gauge invariant and vertical and therefore admits an expression in terms of the local coordinates $t_i$. 
On the patch  $U_S$ we have,
\begin{equation}\label{dt}
\frac{\d{} t_i}{t_i}= \frac{\d{} z_i}{z_i} - \sum_{a\in S} \sum_{b=1}^s \mathcal{Q}^a_b Q_i^b \frac{\d{} z_a}{z_a}~,
\end{equation}
where we took (\ref{local}) into account. 
Setting $i=c\in S$ then gives $\d t_c=0$, {\it cf.} (\ref{covfefe}). This leaves us with $d$ linearly-independent one-forms  $\d t_\alpha$, $\alpha\in {}^\complement{}S$.  
We can then compute,
\begin{equation}\label{Dz}
\frac{\mathcal{D}z_i}{z_i} = \frac{\d{} t_i}{t_i}-h_{ij} |z_j|^2 \frac{\d{} t_j}{t_j}~.
\end{equation}
where we took into account that: 
$h_{ij} |z_j|^2 Q_j^b = \tilde{g}_{cd}\,Q_i^c Q_j^d\, |z_j|^2\,Q_j^b=\tilde{g}_{cd}\,Q_i^c\, g^{db}= Q_i^b$. 
As expected, given that the form on the left-hand side of (\ref{Dz}) is vertical and gauge invariant, the result  can be expressed in terms of the local coordinates alone. 
Note also that the gauge-invariant $|z_j|^2$ can be expressed as a function of  $t_i$  using (\ref{struc}). 

Conversely,  (\ref{dt}) 
can be used to express  $\d{} t_i$ as a function of $\mathcal{D}z_i$, since $\d{} t_i$ is vertical by definition. 
We now have all the necessary tools to translate back and forth between the local coordinate system $\{t_i\}$ on $M$ and the global coordinate system $\{z_i\}$ on $\mathbb{C}^k$.

\subsection{The hermitian metric}

A useful object on an almost complex manifold $M$ is the {\it hermitian metric}, 
\begin{equation}
\mathfrak{h}= \mathfrak{h}^{i\bar{j}} \ \d z_i\otimes \d\bar{z}_{\bar{j}}~,
\end{equation}
where  $\mathfrak{h}^{i\bar{j}}$  can be thought of as a hermitian positive-definite matrix. 
The real and imaginary parts of  $\mathfrak{h}$ are real bilinear forms, so that,
\begin{equation}
\mathfrak{h}=\mathfrak{g}- i J~,
\end{equation}
where $\mathfrak{g}$ is symmetric, positive definite and can be identified with the Riemannian metric, while $J$ 
is antisymmeric and can be identified with the almost symplectic form --which is a  $(1,1)$-form 
with respect to the almost complex structure. In other words the hermitian metric contains both the metric and 
the almost symplectic form of $M$. 

On toric varieties there is a canonical hermitian metric, 
\begin{equation}\label{canme}
\mathfrak{h}(\xi^a)=P\left(\d{} z_i\otimes \d{} \bar{z}_i\right)= \mathcal{D}z_i \otimes\mathcal{D}\bar{z}_i~.
\end{equation}
As we have already noted,  the $\mathcal{D}z_i$ above are not linearly independent and do not form a basis of the cotangent 
bundle of $M$. Using (\ref{Dz}) it is not difficult to see that the hermitian metric takes the following form in local coordinates,
\begin{equation}\label{hermetric}
\mathfrak{h}(\xi^a)
= \frac{|z_i|^2}{|t_i|^2} \d{} t_i \otimes \d{} \bar{t}_i - h_{jk}\frac{|z_j|^2}{|t_j|^2}\frac{|z_k|^2}{|t_k|^2}\bar{t}_j\d{} t_j\otimes t_k \d{} \bar{t}_k~,
\end{equation}
where we have made use of the identity $h_{ij}h_{ik}|z_i|^2 =h_{jk}$ which can be shown by taking into account the various definitions. 

In this case  $J$ is in fact the K\"{a}hler form of the toric manifold,
\begin{equation}
J = \frac{i}{2} \mathcal{D}z_i \wedge\mathcal{D}\bar{z}_i~.
\end{equation}
Although $\mathcal{D}z_i$ are not closed, it can readily be verified that $\d{} J$ vanishes as it should. 

Let us illustrate the above with the example of  $\cp{k-1}$: on the patch  $U_k$ we have $g = \xi$ and $h_{ij} = \frac{1}{\xi}$. Moreover (\ref{struc}) gives 
$|z_k|^2 =\frac{|z_\alpha|^2}{|t_\alpha|^2}= \frac{\xi}{1+t^2}$, where $t^2:=\sum_\alpha |t_\alpha|^2$; $\alpha=1,\dots,k-1$. Hence, 
\begin{equation}\label{fs}
\mathfrak{h}(\xi)=\xi\left(  \frac{\d{} t_i \otimes \d{} \bar{t}_i}{1+t^2}- \eta
\otimes \bar{\eta}\right)~,
\end{equation}
where $\eta :=  \frac{1}{1+t^2} \ \bar{t}_i \d{} t_i$. 
We thus recover the  Fubini-Study metric  and its associated  Kähler form.

A hermitian metric also gives rise to a scalar product ``$\cdot$'' on forms on $M$. Since $P^2=P$, the calculation of the scalar product on vertical forms can be done in the parent space $\mathbb{C}^k$ using the flat metric. Then, using (\ref{Pdz}) we find :
\eq{
\D \bar{z}_i \cdot \D z_j = 2(\delta_{ij}-h_{ij}\bar{z}_i z_j)~,
}
which shows that the $\D z_i$ are not orthogonal.

\subsection{SU(d) structures}\label{sec:sud}

A Riemannian $2d$-dimensional  manifold $M$ with metric $\mathfrak{g}$ and associated ($\mathfrak{g}$-compatible) 
almost complex structure ${I}$ admits a reduction 
of its structure group to $U(d)$. 
At each point over $M$, 
the almost complex structure ${I}$, which need {\it not} be integrable, splits the cotangent space of $M$ into a holomorphic and an antiholomorphic subspace,  
corresponding to the spaces of (1,0)-forms and (0,1)-forms with respect to ${I}$.  Furthermore a holomorphic top form 
can be defined, {\it i.e.} a $(d,0)$-form with respect to ${I}$, which transforms as a section of the canonical bundle of ${I}$.  The 
canonical bundle of ${I}$ is trivial, and so has vanishing Chern class: $c_1({I})=0$, precisely when it has a non vanishing global section, {\it i.e.} when there is a nowhere-vanishing holomorphic top form. In that case the structure of the manifold is further 
reduced to $SU(d)$. An 
equivalent description of an $SU(d)$ structure on $M$ is given by
 a complex decomposable $d$-form  $\Omega$ 
and  a real two-form $J$ such that,
\eq{\label{stru3}
\Omega\wedge J=0~;~~~\displaystyle\frac{i^{d(d+2)}}{2^d}\Omega\wedge\Omega^* = \frac{1}{d!}J^d~.
}
In this formulation the Riemannian metric on $M$ is constructed from the pair $(J,\Omega)$. 

In six real dimensions $(d=3)$ it is well-known that the topological obstruction for the existence of an $SU(3)$ structure is that the 
manifold should be spin.  We can make contact with the discussion of the previous paragraph by noting that $c_1({I})$ modulo 2 is a topological 
invariant,  and $c_1({I})$ is even in cohomology iff $M$ is spin.\footnote{Note however that $c_1({I})$ itself is not a topological invariant. 
A well-known counterexample is $\mathbb{CP}^3$ which admits both a non-integrable almost complex structure (with $c_1({I})=0$) and an integrable one (with $c_1({I})\neq0$). In both cases $c_1({I})$ is even in cohomology, as of course it should, since $\mathbb{CP}^3$ is spin.} Moreover 
 the {\it torsion classes} characterizing the  $SU(3)$ structure are given by the decomposition of $(\d J,\d\Omega)$,
\eq{\spl{\label{ts3}
\d J &= \frac{3}{2}\ \Im ({W}_1 \Omega^*)+{W}_4\wedge J + {W}_3\\
\d\Omega &= {W}_1\ J\wedge J + {W}_2\wedge J + {W}_5 \wedge \Omega~,
}}
where $ {W}_1$ is a function, $ {W}_2$ is a  $(1,1)$-form, $ {W}_3$ is a real $(2,1)\oplus(1,2)$-form, $ {W}_4$ is a real one-form and  
${W}_5$ is a $(1,0)$-form.

As follows from  the previous discussion, it is not always possible to construct an  $SU(d)$ structure on an arbitrary toric variety $M$. 
Although $\mathbb{C}^k$ has a canonical  $SU(k)$ structure given by,
\begin{eqnarray}
\tilde{J} &=& \frac{i}{2} \d{} z_i\wedge\d{}\bar{z}_i\\
\tilde{\Omega} &=& \bigwedge_i \d{} z_i,
\end{eqnarray}
$\tilde{\Omega}$ does not in general descend to $M$. One can always define, 
\eq{\hat{J}:= P(\tilde{J})~,}
which is vertical and gauge invariant:  
 it is the almost symplectic form associated with the  hermitian metric (\ref{hermetric}). However $P(\tilde{\Omega})$ vanishes trivially since there are no 
  $(k,0)$ forms on $M$. 
To obtain a   $(d:=k-s,0)$-form on  $M$
we must contract $\tilde{\Omega}$ with each of the $V^a$ vectors, so that,
\begin{equation}\label{229}
\hat{\Omega} := \frac{1}{\sqrt{\det g}} \ \prod_{a} \iota_{V^a} \tilde{\Omega}~.
\end{equation}
But then  $\hat{\Omega}$ has the same charge as $\tilde{\Omega}$, {\it i.e.} $q^a= \sum_i Q_i^a$, and so it is not gauge invariant.

On the other hand  the pair 
$({\hat{J}},\hat{\Omega})$ does satisfy the compatibility equations (\ref{stru3}), thus defining a {\it local} $SU(d)$ structure on $M$. 
Moreover $\hat{\Omega}$ admits a simple expression in terms of local coordinates\footnote{Thus defined, $\hat{\Omega}$ is compatible with the transition functions, but the $z_i$ are not strictly functions on $U_S$ since they are not gauge-invariant.  A local form could be constructed by substituting $z_i$ with $|z_i|$, at the cost of losing the compatibility with the transition functions.} on $U_S$. 
After some straightforward manipulations we obtain,
\begin{equation}\label{230}
\hat{\Omega}= (-1)^{S} Q_S\,\dfrac{\prod_i z_i}{\sqrt{\det g}} \ \bigwedge_\alpha \dfrac{\d{} t_\alpha}{t_\alpha}~,
\end{equation}
where $a\in S$, $\alpha \in {}^\complement{}S$ and we have defined,
\eq{\label{232}(-1)^S := (-1)^{\sum_{a\in S} a +\frac{(s+1)(s+2)}{2}}~.}
%
%
%
In \cite{Larfors:2010wb} a prescription was given for the  construction of  {\it global} $SU(d)$ structures on $M$.\footnote{Originally presented for $d=3$, the presciption of \cite{Larfors:2010wb} is in fact  directly generalizable to any dimension.} It relies on the existence of a one-form $K$ on $\mathbb{C}^k$ with the following properties:  
\begin{enumerate}
\item It is vertical and (1,0) with respect to the  
complex structure of $\mathbb{C}^k$.
\item It has half the charge of $\tilde{\Omega}$.
\item It is nowhere-vanishing.
\end{enumerate}
Given a one-form $K$ on $\mathbb{C}^k$ satisfying the conditions above, \cite{Larfors:2010wb} showed that a global $SU(3)$ structure on $M$ can be constructed, and provided 
explicit examples of such a $K$ for certain toric $\mathbb{CP}^1$ bundles. Many more examples of $K$ were provided for other toric varieties in \cite{Larfors:2013zva}, which also provided explicit computations of the torsion classes of the associated $SU(3)$ structures.  
However there is no known construction for $K$ that would be applicable in general, even for a subclass of SCTV, and the search for $SU(3)$ structures on SCTV had so far proceeded in  a case by case fashion. 

In the following we will present a construction of $SU(3)$ structures valid for toric $\mathbb{CP}^1$ bundles over any 2d SCTV. 
As we will see, our method is not equivalent to the prescription of \cite{Larfors:2010wb}, 
although it also makes use  of a certain 
(1,0)-form on $\mathbb{C}^k$.


\subsection{Toric \boldmath$\mathbb{CP}^1$\unboldmath~{}bundles over SCTV}\label{sec:toricb}

In \cite{Oda:1978},  the 
 classification of SCTV in three (complex) 
dimensions was shown to reduce to the classification of certain weighted triangulations of the two-dimensional 
sphere. In \cite{Larfors:2010wb}  it was shown how to systematically translate the results of \cite{Oda:1978} 
into the symplectic quotient language 
reviewed previously. In the following we will be interested in  the subclass of the classification of \cite{Oda:1978} corresponding to $\mathbb{CP}^1$ bundles over a two-dimensional SCTV base. 
However the formalism applies generally to the case of $\mathbb{CP}^1$ bundles over SCTV, so 
in this subsection we will keep the dimension of the base arbitrary.

The $U(1)$ charges of these bundles are given by the following set of $(s+1)\times(k+2)$ matrices, 
\eq{\label{pp}
Q_{I}^A=\left(\begin{array}{ccc}
q_i^a & -n^a & 0\\
0     & 1    & 1
\end{array}
\right)~,
}
where $A=1,\dots,s+1$, $I=1,\dots, k+2$; 
 $n_a\in\mathbb{N}$, $a=1,\dots,s$, are integers specifying the 
twisting of the $\mathbb{CP}^1$ bundle over a SCTV $M$;  
$q_i^a$, $a=1,\dots,s$, $i=1,\cdots, k$, are the $U(1)$ charges 
of the symplectic quotient description of $M$, 
which  is therefore of complex dimension $d=k-s$. (In subsequent subsections we will specialize to the case 
$d=2$.)

The total space of the bundle is constructed by appending  two coordinates and one new charge to those of $M$ (given by the $q_i^a$),  as in (\ref{pp}), thus obtaining a space of complex dimension $d+1$. We will  use the following notation for the data related to the fiber,  
\eq{\label{uvdef}
u:= z_{k+1} ~; ~~~ v:=z_{k+2}  ~; ~~~ \xi:=\xi^{s+1}~.
}
The  last charge $Q_i^{s+1}$ defines a $\cp{1}$ fiber over $M$, while the integers $n^a$ determine the twisting of the bundle. 
Indeed the moment map equations for the total space read,
\eq{\label{rho}\sum_{i=1}^k q_i^a\,|z_i|^2= \xi^a +n^a |u|^2 ~; ~~~ |u|^2+|v|^2=\xi~.}
Thus the last 
two coordinates define a sphere of radius $\sqrt{\xi}$, while the first $n$ coordinates define locally an $M_\rho$ whose ``radii'' $(\rho^a)^2:= \xi^a +n^a |u|^2$ {depend on the fiber}.  The twisting can be thought of as a consequence of the modified $U(1)^{s+1}$ action.

We would now like to construct a metric that exhibits the bundle structure, \textit{i.e.} a metric of the form  $\mathfrak{h}_{d+1}= \mathfrak{h}_d + \mathfrak{h}_{\cp{1}}$, where 
$\mathfrak{h}_d$ is a metric on $M$ and $\mathfrak{h}_{\cp{1}}$ is a metric on the fiber $\cp{1}$, possibly modified by a connexion on the base. 
 By expanding the canonical metric (\ref{canme}) we find,
\eq{\label{233}
\mathfrak{h}_{d+1}(\xi^A)= \mathfrak{h}_d((\rho^a)^2) +\frac{\hat{g}}{g} |u|^2 |v|^2\, \varepsilon\otimes{\varepsilon^*}~,
}
where the details of the computation, which are somewhat involved, can be found in appendix \ref{app:1} together 
with the definitions of the various quantities in the second term of the right-hand side above. The one-form $\varepsilon$ can be thought of 
as an analogue of the vertical displacement along the fiber.

\section{\boldmath$\mathbb{CP}^1$\unboldmath~{}over \boldmath$\mathbb{CP}^2$\unboldmath}\label{sec:3}

Let us now  examine in detail the construction of 
an $SU(3)$ structure on 
the $\mathbb{CP}^1$ bundle over $\mathbb{CP}^2$. This is the simplest example 
in the class of 3d SCTV of the form $\mathbb{CP}^1$ bundle over $M$, where $M$ is a 2d SCTV, but it already captures the main 
idea of the construction. We will treat the general case in section \ref{sec:general}.

The toric data in this case are:  $k=5$ (the complex dimension of the parent space), $s=2$ (the number of charges), 
$d=3$ (the complex dimension of the toric variety). Explicitly the charges are given by,
\begin{equation}\label{charge}
Q=\left(\begin{array}{ccccc}
1&1&1&-n&0\\
0&0&0&1&1
\end{array}\right)~,
\end{equation}
where $n\in\mathbb{N}$. 
The corresponding moment map equations read, using the notation introduced in section \ref{sec:toricb}, 
\eq{\spl{\label{struc2}
|z_1|^2+|z_2|^2+|z_3|^2 &= \xi_1 +n|u|^2\\
|u|^2+|v|^2 &= \xi
~.}}
This is a  $\cp{1}$ bundle over $\cp{2}$, 
with twisting parameterized by  $n$. We can make this more explicit in local coordinates: on the patch $U_{1,5}:=\{z_1,v\neq0\}$ we define,
\begin{equation*}
t_2 := \dfrac{z_2}{z_1}~;~~~
t_3 := \dfrac{z_3}{z_1}~;~~~
t_4 := \dfrac{z_1^nu}{v}
~.
\end{equation*}
Hence $t_2,t_3$ are local coordinates parameterizing a $\cp{2}$ whereas, for $z_1$ fixed, $t_4$ is a local coordinate on a $\cp{1}$. For $n=0$, the bundle becomes trivial and we obtain the direct product $\cp{2}\times\cp{1}$. 
We can also see explicitly  that the toric variety can be covered with patches of the form $U_S$, as in (\ref{local}): in the present case $S$ is given by the pair $(i,j)$ where $i=1,2,3$ and $j=4,5$, and the moment map equations (\ref{struc2}) exclude the simultaneous vanishing of $z_1,z_2,z_3$  or that of $u,v$. To make contact with our previous discussion of local coordinates, we can check here that $|Q_S|=1$ for all the $S$ defined above. On the other hand for the patch $U_{4,5}$ we do not get  compatible local coordinates in general, since $Q_{S'=\{4,5\}}=-n$, however this patch is not used  in the covering of the toric variety by $U_S$.

Let us now calculate explicitly the various objects introduced in section \ref{sec:formalism}. Since the base is defined by only one charge $q_i=(1,1,1)$, the calculations are rather simple. We have,
\eq{\spl{
\hat{g} &= \rho^2\\
V_i     &= \frac{n}{\rho^2}\\
V       &= \frac{n^2}{\rho^2}\\
g       &= \xi \rho^2 + n^2 |u|^2|v|^2
~.}}
We thus find,
\begin{equation*}
\varepsilon = \frac{\d{} t_4}{t_4}+ n \, \eta
~,
\end{equation*}
where we have set,
\eq{\eta:= \frac{1}{1+t^2}\left(\bar{t}_2 \d t_2+\bar{t}_3 \d t_3\right)~;~~~
t^2:= |t_2|^2+|t_3|^2
~.}
If we now introduce, 
\begin{equation*}
 \Gamma := \frac{|u|^2|v|^2}{g}
~,
\end{equation*}
the decomposition of the metric (\ref{233}) can be written,
\begin{equation}\label{fs2}
\mathfrak{h}= \rho^2 \mathfrak{h}_{\cp{2}} \,+ \Gamma \rho^2 \left| \frac{\d{} t_4}{t_4}+n\, \eta\right|^2~,
\end{equation}
where $\mathfrak{h}_{\cp{2}}$ is the hermitian Fubini-Study metric of ${\cp{2}}$ with unit radius, {\it cf.} eq.~(\ref{fs}).

We see the fibration structure appearing naturally in (\ref{fs2}): the displacement along  $t_4$ 
is modified by a connection, proportional to $\eta$, depending on the variables of the ${\cp{2}}$ base, $t_2,t_3$. Moreover,   
\begin{equation}\label{connexion}
\d{} \eta = 2i \hat{j}~,
\end{equation}
where $\hat{j}$ is the K\"{a}hler form of ${\cp{2}}$, {\it cf.} eq.~(\ref{fs}). 
 For vanishing $n$ the connection piece drops out from the vertical displacement and 
the metric becomes that of a direct product as excpected.

\subsection{Comparison with the literature}\label{sec:lit}

Endowed with the hermitian metric (\ref{fs}), the base $\cp{2}$ of the $\cp{1}$ fibration is a  K\"{a}hler-Einstein manifold obeying, 
\eq{\label{ke}
\d{} \hat{j} = 0~;~~~
R_{mn} = \lambda ~\!\mathfrak{g}_{mn}
~.} 
{\it i.e.} $\hat{j}$ is closed and the  Ricci tensor is proportional to the metric. With our conventions,  setting $\xi=1$ gives $\lambda=6$. Identifying the 
 $\cp{1}$ fiber with $S^2$ (by forgetting the complex structure), $M$ can be thought of as an $S^2$ fibration over a  Kähler-Einstein base $B_4$, denoted by $S^2(B_4)$. 
 These spaces appear naturally  
in the context of supersymmetric AdS$_4$ compactifications of M-theory  on the so-called $Y^{p,q}(B_4)$ spaces \cite{Gauntlett:2004hh,Martelli:2008rt}, which  can be thought of as $S^1$ fibrations over $S^2(B_4)$. Compactifying M-theory on an appropriately chosen $S^1$ then leads to $\mathcal{N}=2$ type IIA solutions of the form  AdS$_4\times S^2(B_4)$ \cite{Petrini:2009ur}. The latter can be deformed to solutions of massive IIA for any Kähler-Einstein base $B_4$ \cite{Lust:2009mb}, although regularity requires $B_4$ to have positive curvature.

In the conventions of \cite{Martelli:2008rt} the $S^2(B_4)$ metric reads,
\begin{equation}\label{y6}
\mathfrak{g} = U^{-1} \d{} \tilde{\rho}^2 + \tilde{\rho}^2 \mathfrak{g}_{\cp{2}}+ q\, \left(\d{} \psi + A\right)^2~,
\end{equation}
where $\tilde{\rho}\in\left[\tilde{\rho}_1,\tilde{\rho}_2\right]$ and $\psi \in\left[0,{2\pi}/{3}\right]$ are the coordinates of the $S^2$ fiber 
(for general $\lambda$ the period of  $\psi$ is ${4\pi}/{\lambda}$); $U$ and $q$ are positive functions of  $\tilde{\rho}$, vanishing at  $\tilde{\rho}_1$ et $\tilde{\rho}_2$.  The circle parameterized by  $\psi$ is fibered over the $\left[\tilde{\rho}_1,\tilde{\rho}_2\right]$ interval. The connection $A$ is a one-form on the base $B_4$ obeying,
\eq{\label{ccon}
\d{} A = 2 \hat{j}
~.
}
At the endpoints of the $\tilde{\rho}$ interval the $\psi$ circle contracts to a point, thus 
 resulting in a total space with the topology of $S^2$. The period of $\psi$ is fixed by requiring  the metric to be  smooth at the endpoints, 
 {\it i.e.} that,   
\eq{U^{-1} \d{} \tilde{\rho}^2 + q\,  \d{} \psi ^2\rightarrow \d u^2+u^2 \d\tilde{\psi}^2~,~~~\mathrm{for}~ \tilde{\rho}\rightarrow \tilde{\rho}_1,\tilde{\rho}_2~,}
 where $u$ is a function of $\tilde{\rho}$ that vanishes at the endpoints $\tilde{\rho}_1,\tilde{\rho}_2$, and we have defined
  an angular variable $\tilde{\psi}:=\lambda\psi/2$ with period $2\pi$. 
  
{}Moreover the $\psi$ coordinate parameterizes an $S^1$ fibration in the canonical bundle of $B_4$. To see this, note that the connection of the canonical bundle of 
a K\"{a}hler-Einstein space with curvature normalized as in (\ref{ke})  obeys, 
\eq{\label{cb}
\d \mathcal{P}=\lambda \hat{j}
~,}
{\it cf.}~appendix  \ref{sec:torsion2}. Comparing with (\ref{ccon}) we see that $\mathcal{P}=\lambda A/2$, and so the vertical displacement along the $S^1$ fiber, 
{\it cf.} the last term in (\ref{y6}), is proportional to $(\d\tilde{\psi}+\mathcal{P})$, as required for the canonical bundle. The fact that $\lambda$ is positive for $\mathbb{CP}^2$ guarantees that the total space of the $S^1$ fibration, written in local coordinates in (\ref{y6}), extends globally to a smooth five-dimensional (squashed) Sasaki-Einstein space.

To make contact with the coordinates of (\ref{fs2}), we must rewrite the $\cp{1}$ fiber coordinate $t_4$ in terms of a pair of real coordinates. 
Using the formulas of section \ref{sec:details} we can rewrite the Riemannian metric $\mathfrak{g}$ and K\"{a}hler form ${J}$ associated with (\ref{fs2}) for $n\neq 0$.  The result reads,
\begin{equation}\label{rcan}
\mathfrak{g} = \frac{1}{n^2\Gamma} \d{} {\rho}^2 +{\rho}^2 \mathfrak{g}_{\cp{2}}+ \Gamma\rho^2 \left(\d{} \varphi + n\Im\eta\right)^2~,
\end{equation}
and,
\begin{equation}\label{kf}
{J} = \rho^2 \hat{j} +  \frac{\rho}{n}~\!\d\rho \wedge  \left(\d{} \varphi + n\Im\eta\right)~,
\end{equation}
%
%
%
where we are using local coordinates on the patch $U_{1,5}$, and $\varphi\in[0,2\pi]$ denotes the phase of $t_4$. The $\cp1$ fiber is 
parameterized by the $(\rho,\varphi)$ coordinates: $\varphi$ parameterizes a circle, fibered over the interval $\rho\in[\rho_1,\rho_2]=[\sqrt{\xi_1},\sqrt{\xi_1+n\xi}]$, whose radius vanishes at the endpoints. Indeed $\Gamma$ vanishes for $u=0$ or $v=0$ which correponds respectively to $\rho=\rho_1$ and $\rho=\rho_2$, following from the moment map equations (\ref{struc2}). Moreover it can be checked that the  metric is smooth there.

{}Furthermore we need to deform
the canonical hermitian metric of the toric variety by introducing two warp factors $F(\rho)$, $G(\rho)$ along the base and fiber respectively, 
\eq{\label{fs3}
\mathfrak{h} = F(\rho) \sum_{i=1}^3 \mathcal{D}z_i\otimes\mathcal{D}\bar{z}_i + G(\rho)\sum_{i=4,5} \mathcal{D}z_i\otimes\mathcal{D}\bar{z}_i ~.
}
It can then be seen that the functions $F(\rho)$, $G(\rho)$ together with a change of variables $\tilde{\rho}=\tilde{\rho}(\rho)$ may be chosen so that the real and imaginary parts of 
(\ref{fs3}) reduce to the metric in (\ref{y6}) and the form $J_+$ of \cite{Martelli:2008rt} respectively, provided we set $n=3$. The details of this exercise can be found in appendix \ref{sec:details}.

The condition $n=3$ is also important for the existence of a globally-defined $SU(3)$ structure. We turn to the construction of this structure in section \ref{sec:spc}. 
Note however that the canonical metric of the SCTV, eq.~(\ref{canme}), is smooth by construction for all $n\in\mathbb{N}$. This can also be verified explicitly by examination of the local form of the metric in terms of the coordinates (\ref{local}) in each patch $U_S$.

\subsection{The SU(3) structure}\label{sec:spc}

In this section we will set $F=G=1$ for simplicity of presentation: the two warp factors $F(\rho)$, $G(\rho)$ discussed in section \ref{sec:lit} can be easily 
reinstated without changing any of the conclusions.

Specializing the formalism of section \ref{sec:sud} to the present example  we obtain a local $SU(3)$ structure $(\hat{J},\hat{\Omega})$, where $\hat{J}$ 
is obtained from (\ref{kf}) by setting $n=3$. 
On the other hand we have,
\begin{equation}
\hat{\Omega}= -\dfrac{z_5^2}{\sqrt{\det g}} \ \d{} t_2 \wedge \d{} t_3 \wedge \d{} t_4~,
\end{equation}
which is not gauge invariant, so this $SU(3)$ structure is not globally defined. 
In fact neither of the two local $SU(3)$ structures $(J_\pm,\Omega_{\pm})$ of \cite{Martelli:2008rt} can be globally extended: in the following we will see how to make contact with their results.

Let us first define a local $SU(2)$  structure  $(\hat{j},\omega)$ on $\cp{2}$, where,
\begin{equation}
\omega = \dfrac{1}{(1+t^2)^{3/2}} \ \d{} t_2\wedge \d{}t_3~,
\end{equation}
and $\hat{j}$ is the K\"{a}hler form of ${\cp{2}}$, {\it cf.} eq.~(\ref{fs}), so that,
\begin{equation}\label{su2}
\begin{array}{rcl}
\omega\wedge \hat{j} &=& 0\\
\omega\wedge \omega^* &=& 2 \hat{j}\wedge \hat{j}~.
\end{array}
\end{equation}
This $SU(2)$ structure is only  locally defined since $\omega$ has a  singularity at $z_1=0$, as can be seen by using the transition functions to rewrite $\omega$ in a patch where $z_1$ is allowed to vanish. The  $SU(3)$ structures of \cite{Martelli:2008rt} are then obtained by appending the contribution of the fiber coordinate,
\eq{
J_{\pm} := \rho ^2 \hat{j} \pm\frac{i}{2} K\wedge K^*~;~~~
\Omega_+ := \rho^2\ \omega\wedge K~;~~~
\Omega_- := \rho^2\ \omega\wedge K^*
~,}
where,
\begin{equation}
K:= \rho \sqrt{\Gamma}\ \varepsilon~.
\end{equation}
We see that exchanging $K\leftrightarrow K^*$ is equivalent to $(J_+,\Omega_{+})\leftrightarrow(J_-,\Omega_{-})$.

To better understand  the global properties of the $\Omega_\pm$, let us start from their local expression on the patch $U_{1,5}$,
\begin{eqnarray*}
\Omega_+ &=& e^{-i\varphi} \frac{|z_5|^2}{\sqrt{\det g}} \ \d{} t_2 \wedge \d{} t_3 \wedge \d{} t_4\\
\Omega_- &=& e^{i\varphi}  \frac{|z_5|^2}{\sqrt{\det g}} \ \d{} t_2 \wedge \d{} t_3 \wedge \left(\d{} \bar{t_4} + 3 \bar{t_4} \bar{\eta}\right)~.
\end{eqnarray*}
We can see that the singularity in $\omega$ has been compensated by wedging with $K$, $K^*$. On the other hand, 
we can rewrite $\Omega_\pm$ in the patch $U_{1,4}$ by using the transition function $t_5 = {1}/{t_4}$,
\begin{eqnarray*}
\Omega_+ &=& e^{i\varphi}  \frac{|z_1|^6|z_4|^2}{\sqrt{\det g}} \ \d{} t_2 \wedge \d{} t_3 \wedge (-\d{} t_5)\\
\Omega_- &=& e^{-i\varphi} \frac{|z_1|^6|z_4|^2}{\sqrt{\det g}} \ \d{} t_2 \wedge \d{} t_3 \wedge \left(-\d{} \bar{t_5} + 3 \bar{t_5} \bar{\eta}\right)~.
\end{eqnarray*}
We see that  $\Omega_\pm$ has singularities of the form $e^{i\varphi}={t_4}/{|t_4|}= {|t_5|}/{t_5}$ at $t_4=0$ and $t_5=0$: 
indeed the phase of a complex number $z$ is ambiguous at $z=0$. 
It is always possible to soak up one of the two singularities by multiplying or dividing by $e^{i\varphi}$, but never both at the same time. 
Hence $e^{\pm i \varphi}\Omega_\pm$ are well-defined at $t_4=0$ but singular at $t_5=0$, whereas $e^{\mp i \varphi}\Omega_\pm$ are well defined at $t_5=0$ but singular at $t_4=0$. This problem does not arise for $J_\pm$, since $K\wedge K^*$ does not suffer from any phase ambiguities. 

The way out is then to 
construct an $\Omega$  which combines both $e^{\pm i \varphi}\Omega_\pm$ and $e^{\mp i \varphi}\Omega_\pm$. We can take a hint from the supersymmetric $SU(3)$ structure 
of \cite{Lust:2009mb} which we know is globally well-defined. We use a new coordinate $\theta$ instead of $\rho$, defined by $|u|^2=\xi \sin^2\frac{\theta}{2}$. Thus we see that $|v|^2=\xi\cos^2\frac{\theta}{2}$ and $\rho^2=\xi^1+n\,\xi \sin^2\frac{\theta}{2}$, which means that $\theta =0$ or $\pi$ for $\rho=\rho_1$ (corresponding to $t_4=0$) or $\rho=\rho_2$ (corresponding to $t_5=0$), respectively. The idea is then to modify $\omega\rightarrow\hat{\omega}$ by including the problematic phase $e^{i \varphi}$, then define another form $\tilde{\omega}$ with the property  that  $\tilde{\omega}$ varies from $\hat{\omega}$  to $\hat{\omega}^*$ as $\theta$ varies from 0 to $\pi$. More specifically we define,
\eq{\spl{\label{rcl1}
\hat{\omega}&:= e^{i\varphi}\omega\\
\tilde{j} &:= \sin \theta ~\!\Re\hat{\omega} + \cos\theta~\! \hat{j}\\
\tilde{\omega} &:= \cos\theta~\!\Re\hat{\omega} -\sin\theta~\! \hat{j} + i\Im \hat{\omega}~,
}}
so that the $SU(3)$ structure is given by,
\eq{\spl{\label{rcl2}
J &:= \rho^2 \tilde{j}+\frac{i}{2}K\wedge K^*\\
\Omega &:= \rho^2\ \tilde{\omega}\wedge K~.
}}
The relations (\ref{su2}) ensure that  (\ref{stru3}) is satisfied. Moreover at $\theta=0$ we have $\Omega = e^{i\varphi} \Omega_+$, whereas at $\theta=\pi$ we have 
$\Omega = - \left(e^{i\varphi} \Omega_-\right)^*$. 
%
%
The two singularities have thus been regularized and $\Omega$ is globally defined. Thus the pair  $(J,\Omega)$  is a globally-defined structure $SU(3)$ on the manifold.

Let us make one final comment: the prescription of \cite{Larfors:2010wb} for constructing global $SU(3)$ structures, reviewed at the end of section \ref{sec:sud}, gives a form $\Omega$ which is of type (2,1) with respect to the {integrable} complex structure of the toric variety. We see that the prescription used here can never coincide with that 
of \cite{Larfors:2010wb}:  the form $\Omega$ defined in eq.~(\ref{rcl2}) is of mixed type, varying from (3,0) at $\theta=0$ to (1,2) at $\theta=\pi$, with respect to the {integrable} complex structure.

\section{$\cp1$ over general SCTV}\label{sec:general}

We will now show how to construct a globally-defined $SU(3)$ structure on a canonical (defined in eq.~(\ref{chc}) below) $\cp{1}$ bundle over a SCTV of complex dimension $d=2$. 
This is a generalization, to any SCTV base, of our construction of a  globally-defined $SU(3)$ structure on $\cp1$ over $\cp2$, discussed in section \ref{sec:spc}.

As we saw explicitly in the special case of $\cp1$ over $\cp2$, the canonical metric of the SCTV, eq.~(\ref{canme}), is smooth for any twisting of the bundle parameterized 
by $n^a\in\mathbb{N}$. On the other hand the existence of a globally-defined $SU(3)$ structure imposes a topological constraint 
and hence a constraint on the $n^a$, as we explain in the following. This constraint is automatically  satisfied for the canonical $\cp1$ bundle.\footnote{We use the term {\it canonical metric} for the metric (\ref{canme}) of the SCTV, which is defined for all $n^a$, {\it i.e.}~for all topologies. On the other hand we use the term {\it canonical $\cp1$bundle} for the topology 
defined in eq.~(\ref{chc}). Hopefully this will not lead to confusion.}

We start with a $(d+1)$-dimensional toric $\cp{1}$ bundle over a $d$-dimensional base $M$, whose charges were given in (\ref{pp}). 
The $\cp1$ bundle will be called  {\it canonical}  if the charge of $z_{k+1}$, defining the twisting of the bundle, is taken to compensate exactly for the charges of the base, \textit{i.e.}, 
\eq{\label{chc}n^a = \sum_{i=1}^k q_i^a~.}
As emphasized in \cite{Larfors:2013zva},  the topological condition for the existence of an $SU(3)$ structure on the total space of the SCTV is 
that its first Chern class should be even. Condition (\ref{chc}) guarantees that there is no  topological obstruction for the existence of an $SU(3)$ structure. 
This can be seen as follows: 
the first Chern class of the SCTV is given by,
\eq{\label{c1}c_1=\sum_{I=1}^{k+2} D_I~, }
where we have denoted by $D_I$ the divisors corresponding to $\{z_I=0\}$. On the other hand on a toric variety there are as many linearly-independent 
divisors as there are $U(1)$ charges \cite{Denef:2008wq}. In our case 
the fact that the local coordinates defined by $S$ in (\ref{local}) are gauge-invariant is equivalent to 
the linear relations,
\eq{
D_I-\sum_{A\in S}\sum_{B=1}^{s+1}\mathcal{Q}_B^AQ_I^BD_A=0
~.}
Taking the charges (\ref{pp}) into account, and inserting into (\ref{c1}) then leads to,
\eq{
c_1=\sum_{A\in S}\left(\sum_{b=1}^{s}
\mathcal{Q}_b^A(\sum_{i=1}^{k}q_i^b-n^b)
+2\mathcal{Q}_{s+1}^A
\right)D_A~,
}
which, as advertised, is even if the bundle is canonical. More generally, we see that a globally-defined $SU(3)$ structure exists provided $(\sum_{i=1}^{k}q_i^a-n^a)$ are even 
for all $a$ \cite{Larfors:2013zva}. 

We define the usual toric coordinates  and a local $SU(d+1)$ structure $(\hat{J},\hat{\Omega})$ as explained in section \ref{sec:sud}. 
We recall that $\hat{\Omega}$ is not gauge-invariant: for the canonical $\cp1$ bundle it has charge, 
\eq{Q(\hat{\Omega})=\left(\begin{array}{c} 0 \\ \vdots\\0\\2 \end{array}\right)~,}
where we took (\ref{chc}) into account. 

Following the strategy of section \ref{sec:spc} we would like to define the analogue of the local $SU(2)$ structure $(\hat{j},\hat{\omega})$ on the base $M$, 
{\it cf.}~(\ref{rcl1}). 
As in that case we first note that the $\cp1$ fiber distinguishes  a one-form $K$, which we normalize  such that ${K^*}\cdot K=2$,
\eq{K:= \frac{1}{\sqrt{1-h_{k+2\ k+2}|v|^2}} \mathcal{D}v=\sqrt{\frac{g}{\hat{g}}}\,\frac{\D v}{|u|}~.}
Note that $K$ is not globally defined since it is not gauge-invariant. This can be seen explicitly by taking the  $u\rightarrow0$ limit, in which $\mathcal{D}v$ vanishes. Indeed in this limit we have, $$K\sim \sqrt{\frac{g}{\hat{g}}}v\,\frac{\bar{u}}{|u|}\d{} u \sim e^{i(\varphi_v- \varphi_u)}\d{} u~,$$ where $\varphi_u,\varphi_v$ denote the phases of $u,v$. However $K\wedge {K^*}$ does not suffer from any phase ambiguity,  so that,
\eq{
\hat{j} := \hat{J} - \frac{i}{2} K\wedge {K^*}
~,}
is globally well-defined. Furthermore a somewhat tedious calculation which can be found in appendix \ref{app:ted} shows that $\hat{\Omega}$ can be simplified to,
\eq{\label{ho}
\hat{\Omega} = \frac{(-1)^{d}}{\sqrt{\hat{g}}}\left(\sum_{\hat{S}} (-1)^{\hat{S}} q_{\hat{S}}  \prod_{a\in \hat{S}} z_a\ \bigwedge_{\alpha\in ^\complement{} \hat{S}} \mathcal{D} z_\alpha \right) \wedge e^{i\varphi_u} K
~.}
Its contraction with $K$ is given by,
\eq{
\frac{1}{2} K^*\cdot\hat{\Omega}= \frac{e^{i\varphi_u}}{\sqrt{\hat{g}}}\left(\sum_{\hat{S}} (-1)^{\hat{S}} q_{\hat{S}}  \prod_{a\in \hat{S}} z_a\ \bigwedge_{\alpha\in ^\complement{} \hat{S}} \mathcal{D} z_\alpha \right)
~,}
which is not gauge-invariant.  A gauge-invariant local holomorphic form $\hat{\omega}$ on the base can be constructed as follows,
\eq{
\hat{\omega} := \frac{1}{2} e^{-i\varphi_v} {K^*}\cdot \hat{\Omega}
~.
}
Let us now specialize to $d=2$. 
We can apply the procedure of section \ref{sec:spc} and modify the local $SU(2)$ structure $(\hat{j},\hat{\omega})$ in order to construct a global $SU(3)$ structure. 
Since we have $|u|^2+|v|^2=\xi$, we can define a parameter $\theta\in [0,\pi]$ such that  $|u|= \sqrt{\xi_s}~\! \sin\frac{\theta}{2}$ and 
$|v|= \sqrt{\xi_s}~\!  \cos\frac{\theta}{2}$. 
By the same argument as in section \ref{sec:spc}, the $SU(3)$ structure $(J,\Omega)$  given by,
\eq{\spl{\label{main}
J      &:= j +\frac{i}{2} K\wedge{K^*} \\
\Omega &:= \omega\wedge e^{-i\varphi_v} K~,
}}
where,
\begin{eqnarray*}
j      &:=& \sin\theta\, \Re\hat{\omega}+\cos\theta\, \hat{j} \\
\omega &:=& \cos\theta\,\Re\hat{\omega} -\sin\theta\,\hat{j} + i\Im\hat{\omega}~,
\end{eqnarray*}
can be seen to be  globally-defined. Its associated metric is the canonical metric of the SCTV, given in (\ref{canme}), (\ref{233}).  The associated torsion classes will all be nonvanishing in general, {\it cf.}~appendix \ref{sec:moretor} for more details.

This structure could be easily modified by multiplying $(j,\omega)$ and $K$ by functions of the coordinates of the $S^2$ fiber. 
The associated metric will be modified accordingly to,
\eq{\label{233mod}
\mathfrak{h}_{3} = |h|^2~\! \mathfrak{h}_2 +|f|^2~\! \frac{\hat{g}}{g} |u|^2 |v|^2 K\otimes{K^*}~,
}
for some functions of the fiber coordinates, $f$, $h$.  Indeed modifying the local $SU(2)$ structure via $\omega\rightarrow h^2\omega$, 
$j\rightarrow |h|^2\omega$, $K\rightarrow f K$ results in the metric (\ref{233mod}). 
More generally, an orthogonal transformation can be applied on 
the triplet $(j, \Re\omega,\Im\omega)$, without changing the metric $\mathfrak{h}_2$ of the base.

Provided $f$, $h$ are smooth and nowhere-vanishing, the topology of the total space is that of the SCTV $\cp1$ over $M$. 
The metric (\ref{233mod}) is smooth, since it is a smooth deformation of the canonical metric (\ref{233}) of the SCTV. In some cases allowing $f$, $h$ to 
have singularities or zeros can  lead to a smooth metric on a total space of different topology. We will see an example of this phenomenon in section \ref{sec:lt} where an apparently singular metric on $S^2$ over $\cp2$  is in fact the local form of the round metric on $S^6$.

\section{LT structures on $S^2(B_4)$}\label{sec:lt}

We will now show that the sphere bundles of the form $S^2(B_4)$, where $B_4$ is any four-dimensional K\"{a}hler-Einstein space of positive curvature, admit regular globally-defined $SU(3)$ structures of LT type, {\it i.e.} such that all torsion classes vanish except for $W_1$ and $W_2$. This  is the generic type of $SU(3)$ structure that appears in supersymmetric AdS$_4$ compactifications of massive IIA supergravity \cite{Lust:2004ig}. 

Let $\hat{j}$ be the K\"{a}hler form of $B_4$, normalized as in (\ref{keapp}), (\ref{kemore}) with $\lambda=6$, and let $(\hat{j},\hat{\omega})$ be a local $SU(2)$ structure on 
$B_4$ so that,
\eq{\spl{\label{r1}
\hat{\omega}\wedge {\hat{\omega}}^* &= 2 \hat{j}\wedge\hat{j}~;~~~
   \hat{j}\wedge\hat{\omega}= 0~;\\
\d\mathcal{P} = 6\ \hat{j} ~&;~~~
\d \hat{j}      = 0~;~~~
\d \hat{\omega} = i \p \wedge \hat{\omega} ~,}}
where $\mathcal{P}$ is the canonical bundle of $B_4$, {\it cf.}~appendix \ref{sec:torsion2}.  
We define the following $SU(3)$ structure,
\eq{\spl{\label{s3}
J             &=  |h|^2j + \frac{i}{2}K\wedge K^*\\
\Omega        &=  h^2 \omega\wedge K
~,
}}
where $h$ is a complex function of $\theta$ and,
\eq{\spl{\label{r2}
j      &:= \cos \theta\  \hat{j}+\sin\theta\ \Re(e^{i\psi}\hat{\omega})\\
\omega &:= -\sin\theta\ \hat{j}+\cos\theta\ \Re(e^{i\psi}\hat{\omega})+i\ \Im(e^{i\psi}\hat{\omega})\\
K &:= f  \d\theta + i g  (\d\psi +\p)
~,}}
with $\psi\in[0,2\pi)$ and  $f$, $g$ real functions of $\theta$. 
The associated metric  reads,
\eq{\label{am}
\mathfrak{g}  =  |h|^2\mathfrak{g}_{4}+ f^2 \d\theta^2 + g^2 (\d\psi+\p)^2
~,
}
with $\mathfrak{g}_{4}$ the  K\"{a}hler-Einstein metric of $B_4$.

Using eq.~(\ref{r1}), one can then compute the  torsion classes of the $SU(3)$ structure (\ref{s3}),  
\eq{\spl{\label{torsions}
W_1 &= -\frac{i}{3}\frac{h}{{h}^*}\left( \frac{1}{f} + \frac{\sin\theta}{g}+ 6 \frac{g\sin\theta}{|h|^2}\right)\\
W_2 &= \frac{i}{3}\frac{h}{{h}^*}\left( \frac{1}{f} + \frac{\sin\theta}{g}- 12 \frac{g\sin\theta}{|h|^2}\right) J^\perp \\
W_3 &= \frac{1}{2}\left( \frac{1}{f} - \frac{\sin\theta}{g}+ 6 \frac{g\sin\theta}{|h|^2}\right) \Re\Omega^\perp\\
W_4 &= \left(\frac{|h^2|'}{f|h^2|}-6\cos\theta \frac{g}{|h^2|}\right) \d{}\theta\\
W_5 &=  \left( \frac{h'}{fh} + \frac{g'}{2fg}- \frac{\cos\theta}{2g}\right) K
~,}}
where we have introduced the primitive forms,
\eq{\spl{
J^\perp      &= |h|^2\ j - i K\wedge K^*\\
\Omega^\perp &= |h|^2 \ \omega\wedge K^*~.
}}
Moreover, as we show in appendix \ref{sec:torsion}, one can impose $W_3=W_4=W_5=0$ provided,
\eq{\spl{\label{47}
f&=\alpha\left(1-6\alpha^2\frac{\sin^2\theta}{H}\right)^{-1}~;~~~
g=\alpha\sin\theta~;~~~
h=\sqrt{H(\theta)} \ e^{i\beta} ~,
}}
with,
\eq{\spl{\label{48}
H(\theta)&:= \frac{1}{3}\left( \tilde{x}+ \frac{\tilde{x}^2}{B}+B\right)\\
B&:= \left(\frac{27 H_0^3}{2}+\tilde{x}^3+ 3\sqrt{3}\sqrt{\frac{27 H_0^6}{2}+\tilde{x}^3H_0^3}\right)^{1/3}\\
\tilde{x}&:=9\alpha^2 \sin^2\theta
~,}}
where the real constants $\alpha$, $\beta$  and $H_0\geq 0$ are the parameters of the solution. 

For $H_0>0$ the functions $f$, $h$ are nowhere vanishing. Moreover 
the $\theta\rightarrow 0,\pi$ limit gives a regular metric, provided the period of $\psi$ is $2\pi$. Then 
by the same argument as in \cite{Gauntlett:2004hh,Lust:2009mb}, the  $SU(3)$ structure (\ref{s3}) is globally-defined and the associated metric (\ref{am}) is regular and complete: 
the $(\psi,x^{\mu})$ space, where $x^{\mu}$ are the coordinates of $B_4$, parametrizes a circle fibration in the canonical bundle $\mathcal{L}$ over $B_4$; it extends to a complete, regular five-dimensional Sasaki-Einstein manifold provided $B_4$ is K\"{a}hler-Einstein of positive curvature \cite{FriedrichKath}. The $(\psi,\theta)$ space parameterizes a smooth 
$S^2$, so that the total space has the same topology as $\mathcal{L}\times_{U(1)}\cp1$, in the notation of \cite{Gauntlett:2004hh}. The nonvanishing torsion classes read,
\eq{\spl{\label{lts}
W_1 &= -\frac{2i}{3} \frac{e^{2i\beta}}{\alpha}\\
W_2 &= \frac{2i}{3} \frac{e^{2i\beta}}{\alpha} \left(1- \frac{9\alpha^2\sin^2\theta}{H}\right)\ J^\perp
~.}}
Therefore the $S^2(B_4)$ bundles admit $SU(3)$ structures of LT type, rendering them suitable as compactification spaces for supersymmetric AdS$_4$ solutions of massive IIA \cite{Lust:2004ig}. Note that unlike the LT $SU(3)$ structures on $S^2(\cp2)$ discussed in \cite{Tomasiello:2007eq} from the point of view of twistor spaces ({\it cf.}~appendix \ref{sec:twistor}) or in \cite{Koerber:2008rx} from the point of view of cosets, the structure (\ref{lts}) does not obey $\d W_2\in  (3,0)\oplus(0,3)$.\footnote{It should be possible to make contact with the results of 
\cite{Tomasiello:2007eq,Koerber:2008rx} 
 by suitably acting on the vielbein by an orthogonal transformation. There does not seem to exist a simple ansatz for 
 this  transformation, which  may be  rather involved as it could {\it a priori} depend on all coordinates.} 
Indeed a direct calculation gives,
\eq{
\d W_2=e^{2i\beta} \left(1- \frac{9\alpha^2\sin^2\theta}{H}\right)\left( \frac{2i}{3\alpha^2}\left(1- \frac{9\alpha^2\sin^2\theta}{H}\right) \Re(e^{-2i\beta}\Omega) - \frac{6i\sin^2\theta}{H}\,\Re\Omega^\perp\right)
~.
}
As a consequence, if these manifolds are to be used as compactification spaces for massive IIA, the Bianchi identity for the RR two-form will require the introduction of (smeared) six-brane sources. Another difference from the LT structures of \cite{Tomasiello:2007eq,Koerber:2008rx} is that the discussion of this section applies to any $S^2(B_4)$ bundle with K\"{a}hler-Einstein base, not only to $B_4=\cp2$.

In the case $H_0= 0$, on the other hand, one obtains the solution,
\eq{\spl{
f&=3\alpha~;~~~
g=\alpha\sin\theta~;~~~
h=3\alpha\sin\theta \ e^{i\beta} ~.
}}
This corresponds to the {\it nearly K\"{a}hler} limit, in which also $W_2$ vanishes. 
Moreover the  $\theta\rightarrow 0,\pi$ limit results in a conical metric of the form,
\eq{\label{cm}
\mathfrak{g}  \sim  \d\theta^2+\theta^2 \d s^2_{5}
~,
}
where,
\eq{\label{se5}
\d s^2_{5}:=\mathfrak{g}_{4}+ \frac{1}{9} (\d\psi+\p)^2~,
}
is the canonically normalized metric of a five-dimensional Sasaki-Einstein base written as a circle fibration on the canonical bundle over $B_4$; the normalization 
is such that the cone metric (\ref{cm}) is Ricci-flat. Hence for $H_0=0$ the metric presents conical singularities in general, unless $B_4$ is $\cp2$, in which case 
the associated Sasaki-Einstein metric (\ref{se5}) is that of the round sphere, and the associated cone (\ref{cm}) is not only Ricci-flat but also flat. Going back to the 
metric (\ref{am}) we obtain,
\eq{ 
\mathfrak{g}  =  9\alpha^2\left(\d\theta^2+\sin^2\!\theta~\!\d s^2_{5}
\right)
~.
}
We thus see that in the smooth case, $B_4=\cp2$, we obtain a round $S^6$ of radius $3\alpha$. We thus recover the well-know result that the round $S^6$ admits an associated 
nearly-K\"{a}hler structure.

Let us finally note that we may relax the condition on $B_4$, so that $B_4$ is any four-dimensional K\"{a}hler manifold (not necessarily toric, or Einstein). In this case 
the torsion classes can also be explicitly calculated, {\it cf.} appendix \ref{sec:apptorsionkaehler}, however we do not expect the structure to admit a global extension to a complete space with a regular metric.

\section{Conclusions}\label{concl}

The construction of $SU(3)$ structures on SCTV had up to now proceeded on a case-by-case basis. In the present paper we 
gave a formula for a globally-defined $SU(3)$ structure valid on all canonical $\cp1$ bundles over two-dimensional SCTV. 
This $SU(3)$ structure admits a space of deformations parameterized by certain functions, on which 
the associated torsion classes depend. 
The construction  
is genuinely different from that in \cite{Larfors:2010wb}: as 
opposed to the construction in that reference, it produces a holomorphic three-form of varying type (with respect to the 
integrable complex structure of the SCTV).

Having a general  formula for the $SU(3)$ structure opens up the possibility of a systematic (possibly automatized) scan for flux vacua. Such a procedure has been successfully carried out in the case of solvmanifolds \cite{Grana:2006kf} and cosets \cite{Koerber:2008rx}, and would be interesting to undertake also in the 
class $\cp1$ over SCTV  considered here. It could be extended to $\cp1$ fibrations over noncompact toric varieties, as the formalism does not rely on compactness 
other than in the input of the $U(1)$ charges  specifying the toric variety \cite{Chen:2010bn}.

The construction of the $SU(3)$ structure was also applied to the case of $S^2(B_4)$ bundles. These spaces first appeared 
as six-dimensional bases of seven-dimensional Sasaki-Einstein spaces in the context of $\mathcal{N}=2$ AdS$_4$ vacua  of M-theory \cite{Gauntlett:2004hh}. It was 
subsequently realized \cite{Martelli:2008rt} that reducing along the so-called $\alpha$-circle produces a (warped) $\mathcal{N}=2$  AdS$_4\times S^2(B_4)$ vacuum of IIA. 
The relevant supersymmetric $SU(3)$ structure, whose existence was implicitly inferred in  \cite{Martelli:2008rt}, was first constructed explicitly in \cite{Petrini:2009ur} for the 
case $B_4=\cp2$. The generalization to arbitrary $B_4$ was given in \cite{Lust:2009mb}. 

In the present paper we showed that the $S^2(B_4)$ spaces also admit a different $SU(3)$ structure of LT type, thus making them suitable for $\mathcal{N}=1$ 
compactifications of massive IIA. However, these solutions require (smeared) six-brane sources, complicating their physical interpretation. It is possible that a more general 
orthogonal rotation of the 
local $SU(2)$ structure acting on the triplet $(j, \Re\omega,\Im\omega)$ may produce a sourceless LT structure, although the analysis becomes rather cumbersome in this case 
and we have been unable to obtain a conclusive result.\footnote{If the existence of a sourceless LT structure could be established within the context of the ansatz of the present paper,
it would exist for all $S^2(B_4)$ spaces, not only for $B_4=\cp2$, as our ansatz only relies on the K\"{a}hler-Einstein property of the base. 
As already remarked, a sourceless LT structure does exist on $S^2(\cp2)$ \cite{Tomasiello:2007eq,Koerber:2008rx}, but seems to rely on the special properties of 
$\cp2$ as a selfdual Einstein manifold \cite{xu}. This is not taken into account by our ansatz, and would not be applicable to the other K\"{a}hler-Einstein bases $B_4$: $\cp2$ is 
the only K\"{a}hler-Einstein four-manifold of positive curvature that is also selfdual.}




\appendix

\section{Decomposition of the metric\label{app:1}}

In this section we fill in some of the details leading up to eq.~(\ref{233}). 
We start by defining the vertical one-forms using the formalism introduced in section \ref{sec:diff}, for the total bundle. The various objects are thus given 
in terms of the charges $Q_I^A$.  Expressing them in terms of the charges of the base $q_i^a$ we have,
\[\begin{array}{rcl}
\hat{g}^{ab} &:=& q_i^a q_i^b |z_i|^2\\
\hat{h}_{ij} &:=& \hat{g}_{ab} q_i^a q_j^b\\
\hat{\mathcal{D}}z_i &:=& \hat{P}(\d{} z_i) = \d{} z_i - \hat{h}_{ij} z_i \bar{z}_j \d{} z_j~,
\end{array}\]
where hatted symbols are used to denote objects relative to the base, in order to distinguish from 
the objects constructed in (\ref{sec:diff}). Note that $\hat{g},\hat{h},\hat{\mathcal{D}}z_i$ live on the same space as their non-hatted counterparts, which are the relevant objects for the definition of forms in the symplectic quotient description. This means that they do not have any a priori interpretation as objects on the base. For example, the $|z_i|^2$ do not verify the moment map equations of the base but those of the total bundle, and thus  $\hat{g},\hat{h}$ and $\hat{\mathcal{D}}z_i$ depend on the radii.
A quick calculation confirms that the $\hat{\mathcal{D}}z_i$ do obey the expected algebraic relations,
\[\sum_{i=1}^kq_i^a \bar{z}_i \hat{\mathcal{D}}z_i=0~.\]
Recall the form of the canonical metric on a SCTV (the generalization of the Fubiny-Study metric of $\mathbb{CP}^1$),
\[\mathfrak{h}_{d+1} = \sum_{I=1}^{k+2} \D z_I\otimes\D\bar{z}_I~.\]
We will now decompose this metric into base and fiber components. Since the $\D z$ depends on the matrix $g^{AB}$, the key here will be to decompose it and its inverse along the different bundle directions.

The definition of $Q_I^A$ leads to :
\[g^{AB}=\left(\begin{array}{cc}
\hat{g}^{ab} + n^a n^b |u|^2  & -n^a |u|^2 \\
-n^b |u|^2                    &  \xi  
\end{array}\right)~.\]
Moreover we need to express the inverse $g_{AB}$ while keeping track of the inverse, $\hat{g}_{ab}$, of $\hat{g}^{ab}$. 
For this purpose  we first need to compute the determinant $g= \det g^{AB}$,
\[g=\left|\begin{array}{cc}
\hat{g}+ n\, n^T |u|^2 & n |u|^2\\
n^T |u|^2           & \xi
\end{array}\right|
=
\left|\begin{array}{cc}
\hat{g}+ n\, n^T |u|^2(1-\frac{|u|^2}{\xi}) & 0\\
n^T |u|^2           & \xi
\end{array}\right|
=
\xi \det \left(\hat{g}+ \frac{1}{\xi} |u|^2|v|^2 n\,n^T\right)~.
\]
We now use the property of multilinearity of the determinant to expand this expression. We then get all different terms of order $s-m$ in $g$ and $m$ in $n\,n^T$. But since $\rank n\,n^T=1$, only the terms of order zero  or one remain. The terms of order one are merely the determinant of $\hat{g}$ where the column $a$ has been replaced by the vector $\frac{n^a}{\xi}|u|^2|v|^2 \ n$. By expanding along this same column, we exhibit the cofactors of $\hat{g}$ which are independent of this exact column, and are related to the inverse matrix,
\[\det(\hat{g}, g^a \leftrightarrow n) = \sum_a \mathrm{cof}(\hat{g})_{ab} n^b = \hat{g}\,\hat{g}_{ab}n^b~.\]
Thus we have :
\[g = \xi\left(\hat{g} + \frac{n^a}{\xi} |u|^2|v|^2 \hat{g} \,\hat{g}_{ab}n^b  \right)= \hat{g}(\xi + \hat{g}_{ab}n^a n^b |u|^2|v|^2)~.\]
The same trick can be used to compute the inverse matrix :
\[g_{s+1\, s+1}=\frac{1}{g}\det(\hat{g}+n\,n^T |u|^2) = \frac{\hat{g}}{g}(1+\hat{g}_{ab}n^an^b |u|^2)~.\]
Moreover,
\[g_{a\, s+1} = \frac{1}{g}\det(\hat{g},g^a\leftrightarrow -|u|^2\, n)=\frac{\hat{g}}{g}|u|^2\, \hat{g}_{ab}n^b~.\]
The last cofactors are somewhat more complicated, since they involve double cofactors. Eventually we get :
\[g_{ab} = \hat{g}_{ab}-\frac{\hat{g}}{g}|u|^2|v|^2\,\hat{g}_{ac}n^c\,\hat{g}_{bd}n^d ~.\]
It is now possible to compute the $h_{\mu\nu}$. Let us introduce the objects 
\eq{\label{vdef}V:= \hat{g}_{ab}n^a n^b~, ~~~V_i:=\hat{g}_{ac}q_i^a n^c~,}
in terms of which we obtain,
\eq{\spl{
h_{ij}&= g_{ab} q^a_i q^b_j = \hat{h}_{ij}- \frac{\hat{g}}{g}|u|^2|v|^2\,V_i\,V_j\\
h_{ik+1} &= g_{a\, s+1}q_i^a-g_{ab} q_i^a n^b = -\frac{\hat{g}}{g} V_i |v|^2\\
h_{i k+2} &= g_{a\,s+1} q_i^a = \frac{\hat{g}}{g}V_i |u|^2\\
h_{k+1\,k+1}&=g_{s+1\,s+1}- 2g_{a s+1}n^a + g_{ab}n^an^b= \frac{\hat{g}}{g}(1+V|v|^2)\\
h_{k+1\,k+2}&= g_{s+1\,s+1} - g_{a s+1}n^a = \frac{\hat{g}}{g}\\
h_{k+2\,k+2}&= g_{s+1\,s+1} = \frac{\hat{g}}{g}(1+V |u|^2)
~.}}
We can now 
compute the $\D z_I$,
\eq{\label{DzK}
\frac{\D z_i}{z_i}= \frac{\hat{\D}z_i}{z_i}+\frac{\hat{g}}{g}V_i |u|^2|v|^2 \ \varepsilon ~,}
where,
\eq{\varepsilon = \frac{\d{} u}{u}-\frac{\d{} v}{v}+ V_j \bar{z}_j\d{} z_j~.} 
The last two coordinates correspond to colinear one-forms,
\[\frac{\D u}{u}= \frac{\hat{g}}{g}|v|^2\ \varepsilon;~~~\frac{\D v}{v}= -\frac{\hat{g}}{g}|u|^2\ \varepsilon~.\]
Finally the canonical metric reads,
\eq{\spl{
\mathfrak{h}_{d+1} &= \D z_i \otimes \D \bar{z}_i +\D u\otimes\D\bar{u}+\D v\otimes\D\bar{v}\\
				   &= \hat{\D} z_i \otimes \hat{\D} \bar{z}_i + \frac{\hat{g}}{g}V_i |u|^2 |v|^2\ \hat{\D}z_i \otimes \bar{z}_i {K^*} +c.c \\
				  &~~~~~~~~~~~~~~~~~~+ V_iV_i |z_i|^2 \frac{\hat{g^2}}{g^2}|u|^4 |v|^4 \ \varepsilon\otimes{\varepsilon^*}+ \frac{\hat{g}^2}{g^2} |u|^2 |v|^2 \xi \ \varepsilon\otimes {\varepsilon^*}~.
}}
On the other hand we have, 
\[V_i \bar{z}_i \hat{\D}z_i = \hat{g}_{ab} n^a\, q_i^b \bar{z}_i \hat{\D}z_i = 0~;~~~
V_i^2\, |z_i|^2 = V~,\]
so that the metric simplifies to,
\begin{eqnarray}\label{decomp}
\mathfrak{h}_{d+1}(\xi^A) &=&  \mathfrak{h}_d((\rho^a)^2) +  \frac{\hat{g^2}}{g^2} |u|^2 |v|^2 (\xi + V|u|^2|v|^2) \ \varepsilon\otimes {\varepsilon^*}\\
				   &=& \mathfrak{h}_d((\rho^a)^2) +\frac{\hat{g}}{g} |u|^2 |v|^2 \ \varepsilon\otimes {\varepsilon^*}~.
\end{eqnarray}
Note that this decomposition remains valid in the complex local coordinates $t_i,t_{k+1}$, on the chart $U_S$ defined by $S=\hat{S}\cup\{k+2\}$, in which $\varepsilon$ can be written as,
\[\varepsilon = \frac{\d{} t_{k+1}}{t_{k+1}}+ \sum_{i=1}^k V_i |z_i|^2 \frac{\d{} t_i}{t_i}~.\]
The $\hat{\D}z_i$ happen to be the projections on the space generated by the $\d{} t_i$, in fact they are related to the $\d{} t_i$ by the relations (\ref{Dz}) where we take $\hat{h}_{ij}$ instead of $h_{IJ}$. This justifies that in the decomposition (\ref{decomp}), the metric on the base is exactly the canonical metric whose radii vary along the fiber.


\section{K\"{a}hler-Einstein manifolds}\label{sec:torsion2}


 A K\"{a}hler manifold of real dimension $2d$ corresponds to the case of a {\it local} $SU(d)$ structure where $W_5$ is the only 
nonvanishing torsion class,  
\eq{ 
\d J=0~;~~~
\d\Omega= i\mathcal{P}\wedge\Omega ~,
\label{kaehlerclassesb}
}
{\it cf.}~(\ref{ts3}). 
The local structure $(J,\Omega)$ can also be expressed in terms of  bilinears of  a locally-defined spinor $\zeta$ on $M$. 
In terms of this spinor  eq.~(\ref{kaehlerclassesb}) can be written equivalently, 
\eq{
\nabla_m\zeta=\frac{i}{2}\mathcal{P}_m\zeta~,
\label{kaehlerclasses}
}
where  $\mathcal{P}:=2\Im W_5$ is a real one-form. 
(Note that the existence of the complex structure allows us to reconstruct the torsion $W_5$ from its imaginary part alone.) 
%
%
Moreover (\ref{kaehlerclasses}) can be inverted to obtain $\mathcal{P}$ from the 
covariant spinor derivative,
\eq{\label{1}\mathcal{P}_m=-2i\zeta^{\dagger}\nabla_m\zeta~.}
From (\ref{1}),(\ref{kaehlerclasses}), using $\nabla_{[m}\nabla_{n]}\zeta=\frac18R_{mnpq}\gamma^{pq}\zeta$ we obtain,
\eq{\label{d5}
\d \mathcal{P}=\mathcal{R}~,}
where $\mathcal{R}$ is the Ricci form. 
Hence $\mathcal{P}$ can be identified with the connection of the canonical bundle of $M$. 
On the other hand, the Ricci tensor is obtained from 
the Riemann tensor via,
\eq{\mathcal{R}_{mn}=\frac12 R_{mnpq}J^{pq}= R_{mpnq}J^{pq}~.}
On a K\"{a}hler manifold the Ricci form, the Ricci tensor and the Ricci scalar obey,
\eq{\label{rf}\mathcal{R}_{mn}=J_m{}^pR_{pn}~;~~~
\mathcal{R}_{mn}J^{mn}=R~.}
Furthermore for a K\"{a}hler-Einstein manifold such that, 
\eq{\label{keapp}R_{mn}=\lambda g_{mn}~,}
eqs.~(\ref{keapp}),(\ref{rf}) imply,
\eq{\label{kemore}\mathcal{R}=\lambda J~,}
but in general the Ricci form need {\it not} be proportional to the  K\"{a}hler form.

The above relations are valid for arbitrary dimension. 
Specializing to four real  dimensions we adopt the notation $(J,\Omega)\rightarrow(\hat{j},\hat{\omega})$, in accordance with the 
main text. 
We may decompose any two-form $\Phi$ on the basis of a local $SU(2)$ structure $(\hat{j},\hat{\omega})$ as follows:
\eq{\Phi=\varphi \hat{j}+\widetilde{\Phi}+\chi\hat{\omega}+\psi\hat{\omega}^*~,}
where $\varphi:=\frac14 \hat{j}^{mn}\Phi_{mn}$ is the trace of $\Phi$, and $\widetilde{\Phi}$ is (1,1)-traceless: $\hat{j}^{mn}\widetilde{\Phi}_{mn}=0$. Equivalently,
\eq{\hat{j}\wedge\widetilde{\Phi}=0~.}
It is also straightforward to show that $(\hat{j},\hat{\omega})$ are selfdual forms while 
(1,1)-traceless forms are anti-selfdual,
\eq{\star(\hat{j},\hat{\omega})=(\hat{j},\hat{\omega})~;~~~\star\widetilde{\Phi}=-\widetilde{\Phi}~.}
In particular for the Ricci form the expansion reads,
\eq{
\mathcal{R} = \frac14 R \hat{j}+\tilde{R}
~.}
Moreover the above properties can be used to calculate,
\eq{\mathcal{R}\wedge\mathcal{R}=\left(
\frac14 R^2-\frac12 R_{mn}R^{mn}
\right)\mathrm{vol}_4~,}
where the volume is given by,
\eq{\mathrm{vol}_4=\frac12 \hat{j}\wedge \hat{j}~.}
%


\section{Real coordinates}\label{sec:details}

In this section we explain in detail how to rewrite the hermitian metric (\ref{fs2}) in terms of real coordinates, and make contact with 
the metric (\ref{y6}). 
Let us start by rewriting the $\cp{1}$ fiber coordinate $t_4$ in terms of a pair of real coordinates. 
It is not necessary to do the same for 
 $t_2,t_3$, since the coordinates of the  $\cp{2}$ base do not appear 
 explicitly in (\ref{y6}). Using  eq.~(\ref{rho}), $|t_4|$ can be written in terms of  $\rho$ and the base coordinates,
\begin{equation}
|t_4|^2 = \frac{|z_1|^{2n}|z_4|^2}{|z_5|^2}=\frac{\rho^{2n}}{(1+t^2)^n}\frac{\rho^2-\rho_1^2}{\rho_2^2-\rho^2}~.
\end{equation}
Let $\varphi\in [0,2\pi]$ be the phase  of $t_4$, so that $t_4$ becomes a function of  $t_2,t_3,\rho,\varphi$,  
\begin{eqnarray*}
\frac{\d{} t_4}{t_4} &=& \frac{\rho \d{} \rho}{\rho^2-\rho_1^2}+ \frac{\rho \d{} \rho}{\rho^2-\rho_1^2} + n \,\frac{\d{}\rho}{\rho}-n\,\frac{ \d{} (t^2)}{2(1+t^2)}+i\d{}\varphi \\
                    &=&  \frac{\d{}\rho}{n\;\rho\Gamma}-n\Re \eta+i\d{}\varphi~.
\end{eqnarray*}
Moreover we set,
\begin{equation}
\varepsilon:= \frac{\d{} t_4}{t_4}+n\, \eta = \frac{\d{}\rho}{n\,\rho\Gamma} +i\,\left(\d{}\varphi+n\, \Im\eta\right)~.
\end{equation}
The term $|\varepsilon|^2:= \varepsilon\otimes \bar{\varepsilon}$ appears naturally in (\ref{fs2}) through the contribution,
\begin{equation}
\varepsilon\otimes \bar{\varepsilon}= \frac{1}{n^2\rho^2\Gamma^2}\d{}\rho^2+\left(\d{}\varphi+n\, \Im\eta\right)^2 - i \frac{1}{n\,\rho\Gamma}\ \d{}\rho\wedge\left(\d{}\varphi+n\, \Im\eta\right)~.
\end{equation}
The last  term on the right-hand side above contributes to the K\"{a}hler form, while the rest contributes to the metric. Setting $\psi :={\varphi}/{n}$ and $A := \Im\eta$, we recover the terms appearing in (\ref{y6}), provided we set $n=3$. Moreover the relative coefficient between the $\d{}\rho^2$ and the $(\d{}\psi + A)^2$ term is fixed in the expression of  
$|\varepsilon|^2$, and this determines the change of variables $\rho \rightarrow \tilde{\rho}(\rho)$ by comparing with (\ref{y6}). 
However, performing this change of variables in (\ref{fs2}) does not directly bring us to the metric of (\ref{y6}): there remain two coefficients that still need to be adjusted. 
This can be achieved by introducing the two warp factors  of eq.~(\ref{fs3}) as we now show.

Let us go back to the expression of the metric in terms of $\mathcal{D}z_i$. In local coordinates we have,
\eq{\spl{
\dfrac{\mathcal{D}z_1}{z_1} &= n\,\Gamma\varepsilon-\eta \\
\dfrac{\mathcal{D}z_2}{z_2} &= \dfrac{\d{} t_2}{t_2}+n\,\Gamma\varepsilon-\eta \\
\dfrac{\mathcal{D}z_3}{z_3} &= \dfrac{\d{} t_3}{t_3}+n\,\Gamma\varepsilon-\eta \\
\dfrac{\mathcal{D}z_4}{z_4} &= \rho^2\dfrac{\rho_2^2-\rho^2}{n\,\det g}\ \varepsilon \\
\dfrac{\mathcal{D}z_5}{z_5} &= \rho^2\dfrac{\rho_1^2-\rho^2}{n\,\det g}\ \varepsilon  ~.
}}
It follows that the term $\sum_{i=1}^3 \mathcal{D}z_i\otimes\mathcal{D}\bar{z}_i$ gives the hermitian metric of $\cp{2}$ plus a  $|\varepsilon|^2$ term, 
whereas $\mathcal{D}z_4,\mathcal{D}z_5$ only  contribute to $|\varepsilon|^2$. Let us  define,
\eq{\spl{
\mathfrak{h} &= F(\rho) \sum_{i=1}^3 \mathcal{D}z_i\otimes\mathcal{D}\bar{z}_i + G(\rho)\sum_{i=4,5} \mathcal{D}z_i\otimes\mathcal{D}\bar{z}_i\\
             &= F \rho^2 \,\mathfrak{h}_{cp{2}} + \left( F+(\frac{1}{n^2\Gamma}-1)G\right)n^2\rho^2\Gamma^2 |\varepsilon|^2\\
             &=  F \rho^2 \,\mathfrak{h}_{cp{2}} + \left( F+(\frac{1}{n^2\Gamma}-1)G\right) \left(\d{} \rho^2 +n^4\rho^2\Gamma^2 \left(\d{} \psi + A\right)^2 -i n^2\rho \Gamma \, \d{}\rho \wedge\left(\d{} \psi + A\right)\right)~.
}}
We can then adjust  $F$,$G$, and $\rho$ so that,
\eq{\spl{
F \rho^2 &= \tilde{\rho}^2\\
\left( F+(\frac{1}{n^2\Gamma}-1)G\right) \, \d{} \rho^2 &= \dfrac{1}{U} \d{} \tilde{\rho}^2\\
\left( F+(\frac{1}{n^2\Gamma}-1)G\right)n^4 \rho^2\Gamma^2 &= q
~.
}}
These equations can easily be decoupled by first solving for $\rho$, then for $F$ and finally for $G$. 


\section{General SCTV base}\label{app:ted}

In the following we give the details of the derivation of eq.~(\ref{ho}). 
The first step is writing $\hat{\omega}$ in  terms of $\mathcal{D}z$. However this  exercice is rather involved, since the $\mathcal{D}z$ are not independent and because of the ambiguity in the decomposition of wedge products. 
Our starting point is eq.~(\ref{229}),
\begin{eqnarray*}
\hat{\Omega} 
             = \frac{1}{\sqrt{g}} \bigwedge_A Q_J^A z_J\partial_{z_J}\cdot \bigwedge_I \d{} z_I~.
\end{eqnarray*}
In this expression, we notice that the expansion of the contraction with the horizontal vectors amounts to choosing a set $S$ of $s+1$ integers between 1 and $k+2$, corresponding to the indices of the contracted coordinates. We compute,
\[\hat{\Omega} = \frac{1}{\sqrt{g}}\sum_S (-1)^S Q_S  \prod_{A\in S} z_A\ \bigwedge_{\alpha\in ^\complement{} S} \d{} z_\alpha ~,\]
{\it cf.}~(\ref{232}), 
where  $Q_S$ is the determinant of the submatrix of $Q_I^A$ whose columns are indexed by $S$. Notice that if $S$ contains duplicates, or if it  does not select independent columns, the determinant vanishes. Thus the sum selects only the sets $S$ for which the matrix $Q_A^B$ is invertible. The sign $(-1)^S$ is the signature of the permutation required to put the $s+1$ indices of $S$ in the first position, namely :%
\eq{
(-1)^S = \sigma(S,{}^\complement S)=(-1)^{\sum_{a\in S}+\frac{1}{2}(s+1)(s+2)}~.
}

We would  now like to decompose $\hat{\Omega}$ with respect to the bundle structure. We therefore distinguish four cases:
\begin{enumerate}
\item $S\subset [|1,k|]$
\item $S=\hat{S}\cup\{k+1\}$ where $\hat{S}\subset [|1,k|]$, $\sharp\hat{S}=s-1$
\item $S=\hat{S}\cup\{k+2\}$
\item $S=\check{S}\cup\{k+1,k+2\}$ where $\check{S}\subset [|1,k|]$, $\sharp\check{S}=s-2$
\end{enumerate}
In the first case we get $Q_S=0$, since $\rank q_i^a=d<d+1$. In cases 2 and 3 we can easily see that $Q_S=q_{\hat{S}}:=\det(q_a^b)_{a\in\hat{S}}$, while $(-1)^S=(-1)^{\hat{S}}(-1)^d$ for case 2, and $(-1)^S=(-1)^{\hat{S}}(-1)^{d+1}$ for case 3. We can now write,
\[\hat{\Omega} = \frac{(-1)^{d+1}}{\sqrt{g}}\sum_{\hat{S}} (-1)^{\hat{S}} q_{\hat{S}}  \prod_{a\in \hat{S}} z_a\ \bigwedge_{\alpha\in ^\complement{} \hat{S}} \d{} z_\alpha \ \wedge \left( v\d u-u\d v\right)+\frac{1}{\sqrt{g}}\Sigma_4~,\]
with $\Sigma_4$ to be determined. In  case 4 we get, 
\begin{eqnarray*}
Q_S &=& \det(q_a^b,-n^b)_{a\in\check{S}}\\
    &=& \det(q_a^b,-\sum_{i=1}^k q_i^b)\\
    &=& -\sum_{i=1}^k \det(q_a^b, q_i^b)~.
\end{eqnarray*}
In the sum, if $i\in\check{S}$, the determinant cancels out, leaving only a sum over $^\complement{} \check{S}$, so that,
\[\Sigma_4= -\sum_{\check{S}}\sum_{\beta\in ^\complement{}\check{S}} (-1)^S \det(q_a,q_\beta)_{a\in\check{S}} \prod_{a\in \check{S}} z_a u \,v \bigwedge_{\alpha\in ^\complement{}\check{S}}\d{} z_\alpha~.\]
We are now ready include this sum in the one over the $\hat{S}$, which appears in cases 2 and 3: we just need to make the change of variable $\hat{S}=\check{S}\cup\{\beta\}$. However $\d{} z_\beta$ appears in the product, thus we need to shift it to the last position. At the same time we need to move it to its right place inside $\det(q_a,q_\beta)$ so as to maintain the increasing order of $\hat{S}$. The number of shifts needed to do so is the number of shifts required to bring $\beta$ from its place to the end in 
$^\complement{}\check{S}$ plus the number of shifts to bring it from the end to its place in $\check{S}$; since $^\complement{}\check{S}\cup\check{S}= [|1,k|]$, this is exactly the number of shifts required to bring $\beta$ from its place to the end in $[|1,k|]$, \textit{i.e.} $k-\beta$. The last sign we need to compute is,
\begin{eqnarray*}
(-1)^S &=& (-1)^{\sum_{a\in\check{S}} a + (k+1) +(k+2) - \frac{1}{2}(s+1)(s+2)}\\
       &=& (-1)^{\sum_{a\in\hat{S}} a -\beta + (k+1) +(k+2) - \frac{1}{2}s(s+1)-(s+1)}\\
       &=& -(-1)^{\hat{S}} (-1)^{k-\beta +(d+1)}~.
\end{eqnarray*}
Having expressed everything in terms of $\hat{S}$ and $\beta$, it is now possible to transform the sum $\sum_{\check{S}} \sum_{\beta\in^\complement{}\check{S}}$ in $\sum_{\hat{S}}\sum_{b\in\hat{S}}$,  
\[\Sigma_4= (-1)^{d+1}\sum_{\hat{S}} (-1)^{\hat{S}} q_{\hat{S}} \prod_{a\in \hat{S}} z_a \bigwedge_{\alpha\in ^\complement{}\hat{S}}\d{} z_\alpha \ \wedge\left(u \, v \sum_{b\in \hat{S}}\frac{\d{} z_b}{z_b}\right)~.\]
To get a more symmetrical expression we  can simply complete the sum $\sum {\d{} z_b}/{z_b}$, since the missing terms can be trivially added thanks to the wedge product. The final expression is thus,
\[\hat{\Omega} = \frac{(-1)^{d+1}}{\sqrt{g}}\left(\sum_{\hat{S}} (-1)^{\hat{S}} q_{\hat{S}}  \prod_{a\in \hat{S}} z_a\ \bigwedge_{\alpha\in ^\complement{} \hat{S}} \mathcal{D} z_\alpha \right) \wedge \left( v\mathcal{D}u -u\mathcal{D}v + u\,v \sum_{i=1}^k \frac{\mathcal{D} z_i}{z_i}\right)~.\]
The $\d{} z$ were ultimately replaced by $\mathcal{D}z$ because $\hat{\Omega}$ is vertical. Now recall that the expression (\ref{DzK}) decomposes $\D z_i$ into base and fiber parts. Since the metric decomposes correctly into (\ref{decomp}), the $\hat{\D}z_i$ are orthogonal to $K$. Besides, the fiber part can be shown to cancel out in the first factor, so that the first parenthesis is orthogonal to $K$. Thus we can take the second factor to be proportional to $K$, and the proportionality factor can be found by computing,
\eq{\spl{
{K}^* \cdot&\Big( v\mathcal{D}u -  u\mathcal{D}v + u\,v \sum_{i=1}^k \frac{\mathcal{D} z_i}{z_i}\Big) \\
             &= \frac{2}{\sqrt{1-h_{k+2\ k+2}|v|^2}}\Big(v(0- h_{k+1,k+2}\bar{v}\, u)\\             
             &~~~~~~~~~~~~~~{}-u (1-h_{k+2,k+2}|v|^2)+u\,v\sum_{i=1}^k(0-h_{k+2\ i})  \Big)\\
             &= \frac{2u}{\sqrt{1-h_{k+2\ k+2}|v|^2}}\Big(-1+|v|^2(- h_{k+1,k+2} +h_{k+2,k+2}-\sum_{i=1}^k h_{k+2\ i})\Big)~.
}}
On the other hand,
\begin{eqnarray*}
\sum_{i=1}^k h_{k+2\ i} &=&  g_{AB} Q_{k+2}^A  \sum_{i=1}^k Q_i^{B}\\
                        &=& g_{AB} Q_{k+2}^A (Q_{k+2}^B-Q_{k+1}^B)\\
                        &=& h_{k+2,k+2}-h_{k+2,k+1}~,
\end{eqnarray*}
so that,
\[{K^*} \cdot \Big( v\mathcal{D}u -  u\mathcal{D}v + u\,v \sum_{i=1}^k \frac{\mathcal{D} z_i}{z_i}\Big)= -2\sqrt{\frac{g}{\hat{g}}}\frac{u}{|u|}
~.\]
Hence $\hat{\Omega}$ simplifies  to the expression in (\ref{ho}).


\subsection{Torsion classes}\label{sec:moretor}

For a generic SCTV base all torsion classes are nonvanishing.  
We will not write them down explicitly in this case, as they are rather cumbersome and not particularly illuminating. 
The computation boils down to determining the exterior differentials of $\hat{\omega}$ and $K$. In the following we give 
some details of the calculation. 

In the notation of (\ref{uvdef}), $K$ and $(\hat{j},\hat{\omega})$ can be written as follows,
\eq{ 
K= \sqrt{\frac{g}{\hat{g}}}\, \frac{\mathcal{D}v}{|u|}~;~~~
\hat{j} = \hat{J} - \frac{i}{2} K\wedge K^* ~;~~~
\hat{\omega} = \frac{1}{2} e^{-i\varphi_v} K^*\cdot \hat{\Omega}
~.}
In terms of the $\hat{\mathcal{D}}z_i$, we have,
\[\begin{array}{rcl}
\hat{j} &:=& \frac{i}{2}\hat{\mathcal{D}}z_i\wedge\hat{\mathcal{D}}\bar{z}_i\\
\hat{\omega} &:=& \dfrac{e^{i(\varphi_u-\varphi_v)}}{\sqrt{\hat{g}}} \sum_{\hat{S}}(-1)^{\hat{S}} q_{\hat{S}}  \prod_{a\in \hat{S}} z_a\ \bigwedge_{\alpha\in {}^\complement{}\hat{S}} \hat{\mathcal{D}} z_\alpha~.
\end{array}\]
Up to a phase (required for gauge invariance) this coincides with the canonical local $SU(2)$ structure of the base. In particular 
this implies that $\hat{j}$ is K\"{a}hler at fixed fiber coordinates. The dependence of $\hat{j}$ on the fiber coordinates is such that $\hat{J}$ is K\"{a}hler.

We can also rewrite everything in local complex coordinates on the patch $S=\hat{S}\cup\{k+2\}$ :
\eq{\spl{
K            &=-\sqrt{\frac{\hat{g}}{g}}|u|v \ \left( \frac{\d{} t_{k+1}}{t_{k+1}}+V_i |z_i|^2 \frac{\d{} t_i}{t_i}\right)\\
\hat{\omega} &= \dfrac{e^{i(\psi+ \sum_\alpha \psi_\alpha)}}{\sqrt{\hat{g}}} \, (-1)^{\hat{S}} q_{\hat{S}} \prod_i |z^i| \bigwedge_\alpha \frac{\d{} t^\alpha}{t^\alpha}=f \bigwedge_\alpha \frac{\d{} t^\alpha}{t^\alpha} ~,
}}
where $\psi,\psi_\alpha$ are the phases of $t_{k+1},t_\alpha$. We can now introduce real coordinates $\theta,\psi$ on the fiber with $|u|^2=\xi \sin^2\frac{\theta}{2}$:
\[K = \frac{1}{2}\left(\gamma \d{}\theta + \frac{\xi}{\gamma} \sin\theta\ i(\d{}\psi+ A)\right)~,\]
where $\gamma=\sqrt{\frac{g}{\hat{g}}}=\sqrt{\xi +\frac{1}{2}\xi^2\, V\sin^2\theta}$ and $A = V_i\, |z^i|^2\, \Im \frac{\d{} t^i}{t^i}=V_i\, |z^i|^2\, \d{}\psi_i$.
We also get,
\[\d{} A = \frac{i}{2} V_i\, \hat{\mathcal{D}}z_i\wedge\hat{\mathcal{D}}\bar{z}_i+\frac{i}{4} \sin\theta \d{}\theta\,\wedge V_i^2\,(\bar{z}_i\hat{\mathcal{D}}z_i-z_i\hat{\mathcal{D}}\bar{z}_i)
~.\]

Differentiating $\hat{\omega}$ leads to another one-form,
\[\d{}\hat{\omega}= \frac{\d{} f}{f}\wedge\hat{\omega} = \left( -\frac{\xi}{2}V_i(1-\hat{h}_{ii}|z_i|^2) \sin\theta\d{}\theta +i\d{}\psi+ \sum_j i\d{}\psi_j + \frac{1}{2}(1-\hat{h}_{jj}|z_j|^2)\left(\frac{\hat{\mathcal{D}}\bar{z}^j}{\bar{z}^j}-\frac{\hat{\mathcal{D}}z_j}{z_j}\right)\right)\wedge\hat{\omega}~.\]
Alternatively, in terms of $t^i$,
\[\d{}\hat{\omega}= \left( -\frac{\xi}{2}V_i(1-\hat{h}_{ii}|z_i|^2) \sin\theta\d{}\theta + i\left(\d{}\psi+A+\d{}\psi_i |z_i|^2(h_{ii}-h_{ij}h_{jj}|z_j|^2)\right)\right) \wedge \hat{\omega}~.\]
We can write,
\[A':= A+ \d{}\psi_i |z_i|^2(h_{ii}-h_{ij}h_{jj}|z_j|^2)= A+B~,\]
where $B$ comes from the derivatives of $\hat{g}$  and is nonvanishing in general. For simple bases such as $\cp{2}$ or $\cp{1}\times\cp{1}$, $\hat{g}$ is constant and thus $B$ vanishes. The $\d{}\theta$ term comes from the deformation of the base metric along the direction $\theta$. 
Note also that at fixed $\theta$, $\d{}A'\propto\mathcal{R}$ where $\mathcal{R}$ is the Ricci form of the base, {\it cf}.~(\ref{d5}).



\section{LT structures}\label{sec:torsion}

In this section we fill out the details leading up to eq.~(\ref{48}).  
Plugging the following general  ansatz in the  decomposition of $\d{} J,\d{}\Omega$,
\eq{\spl{
\d{} J     &=  \frac{3\alpha_1}{2} \Im \Omega - \frac{3\alpha_2}{2} \Re \Omega + \alpha_3 \Re K + \alpha_4 \Im K+ \alpha_5 \Re \Omega^\perp+\alpha_6\Im\Omega^\perp\\
\d{}\Omega &= a_1 J\wedge J + a_2 K^*\wedge \Omega +a_3 J^\perp\wedge J
~,
}}
for some real  and complex parameters $\alpha_1,\dots,\alpha_6$ and $a_1,\dots,a_3$ respectively, and using eqs.~(\ref{r1}), (\ref{s3}), (\ref{r2}), we arrive at the torsion 
classes given in (\ref{torsions}). 
Imposing $W_3=W_4=W_5=0$ leads to,
\begin{equation}
\begin{array}{rcrcl}
W_3 &:& \frac{1}{f} - \frac{\sin\theta}{g}+ 6 \frac{g\sin\theta}{|h|^2} &=& 0  \\
W_4 &:& \frac{|h^2|'}{f|h^2|}-6\cos\theta \frac{g}{|h^2|}               &=& 0  \\
W_5 &:& \frac{h'}{fh} + \frac{g'}{2fg}- \frac{\cos\theta}{2g}           &=& 0~.
\end{array} 
\end{equation}
From $W_5-\bar{W}_5$ we see that the phase of  $h$ must be constant but is otherwise unconstrained by the equations, {\it i.e.},
\eq{h=|h|e^{i\beta}~,}
for some real constant $\beta\in[0,2\pi)$. Moreover we 
 set $H:=|h|^2$, for some nonnegative function $H$. Since $\d{}\psi$ is not defined at $\theta=0,\pi$, regularity requires that the coefficient of  $\d{}\psi+A$ should vanish at the poles. 
It is therefore convenient to set $g:=G\sin\theta$ for some function $G$. The equations now read, 
\eq{\spl{
  \frac{1}{f} - \frac{1}{G}+ 6 \frac{G\sin^2\theta}{H} &= 0  \\
  H'-3\sin 2\theta \ Gf               &= 0  \\
  \frac{H'}{H} + \frac{G'}{G}+\cot\theta - \frac{f\cot\theta}{G}           &= 0
~,
}}
where we have assumed that $f$, $h$ are nonvanishing. 
Plugging the first two into the third then implies,
\eq{G = \alpha~,}
for some real constant $\alpha$. The system is then solved as in eq.~(\ref{48}), where $H$ satisfies,
\eq{ 
H'\left(1-6\alpha^2\frac{\sin^2\theta}{H}\right)     = 3\alpha^2\sin 2\theta 
~.
}
We immediately see that $H(\theta)= 9\alpha^2\sin^2\theta$ is a special solution. Moreover the differential equation imposes $H(\pi-\theta)=H(\theta)$. It is thus 
convenient to introduce a new function $\varphi(x)$,  where $x:=\sin^2\theta$ and $H:=9\alpha^2 x \varphi(x)$, in terms of which the equation becomes,
\eq{
\frac{\varphi-\frac{2}{3}}{\varphi-\varphi^2}\ \varphi'=\frac{1}{x}
~.}
Integrating over  $x$ between $X_0$ and $X$ we obtain,
\eq{
\int_{X_0}^X \frac{\varphi-\frac{2}{3}}{\varphi-\varphi^2}\ \varphi'\d{} x = \log\frac{X}{X_0}
~,
}
where  $\varphi_0:=\varphi(X_0)$. 
On the other hand,
\eq{
\frac{\varphi-\frac{2}{3}}{\varphi-\varphi^2}= -\frac{2}{3}\frac{1}{\varphi}-\frac{1}{3}\frac{1}{\varphi-1}
~.}
Since  $\varphi\geq 0$ and  $\varphi-1$, $\varphi_0-1$ have the same sign, we find,
\eq{
 -\frac{2}{3}\log\frac{\varphi}{\varphi_0}-\frac{1}{3}\log\frac{\varphi-1}{\varphi_0-1} = \log\frac{X}{X_0}
~,
}
which leads to,
\eq{
\varphi^2(\varphi-1) = \frac{X_0^3}{X^3}\varphi_0^2(\varphi_0-1)
~.}
Rewriting  the above in terms of $H$ which, contrary to $\varphi$, is necessarily everywhere well-defined, we obtain,
\eq{
H^2(H-9\alpha^2 X) = H_0^2(H_0-9\alpha^2 X_0)=\mathrm{constant}
~.}
We can henceforth assume $X_0=0$ without loss of generality, which leads to,
\eq{H^2(H-9\alpha^2 X) - H_0^3=0~.}
It is easy to see that the above polynomial in $H$ is increasing for negative $H$, until it attains the value $-H_0^3\leq 0$ at $H=0$. It then decreases until  $H= 6\alpha^2 X$, from which point on it becomes increasing. Therefore if we impose  $H_0>0$ the polynomial only vanishes once, for $H>6\alpha^2 X\geq 0$. For $H_0=0$, there are two solutions: $H=0$ (which must be discarded) and the special solution $H=9\alpha^2 X$. We conclude that for any $H_0\geq 0$, there is a unique solution to the differential equation with the boundary conditions  $H(0)=H_0=H(\pi)$; it is given in eq.~(\ref{48}) of the main text.

\section{Twistor spaces}\label{sec:twistor}

There is an alternative description of the total space of the $\cp1$ fibration over $\cp2$ in terms of twistor spaces. More generally, for the purposes of the present 
section we may replace the $\cp2$ base by any four-dimensional K\"{a}hler space $B_4$.

Consider $B_4$ equipped with its canonical complex structure  $\hat{I}$ and a hermitian metric  $\mathfrak{g}$. Let us introduce 
a complex zweibein $z_1,z_2$, so that $\hat{I}z_{k}=i~\! z_{k}$, for $k=1,2$. These forms are of course only locally defined, since $B_4$ is not parallelizable in general. 
We can thus express the metric and the local $SU(2)$ structure on $B_4$ in terms of the complex zweibein,
\[\begin{array}{rcl}
\mathfrak{g} &=& z_1 \bar{z}_1+ z_2 \bar{z}_2  \\
\hat{j}      &=& \frac{i}{2}\left( z_1 \wedge\bar{z}_1+ z_2\wedge \bar{z}_2\right)  \\
\hat{\omega} &=& z_1\wedge z_2  ~.
\end{array}\]
At any one point  $x\in B_4$, $\hat{j}_x,\hat{\omega}_x$ form an $SU(2)$ structure on the tangent space $T_xB_4$. The latter is equipped with a complex structure and a scalar 
product given by $\hat{I}_x$ and $\mathfrak{g}_x$ respectively.  Moreover the relation,
\[\hat{I}_m^k = \mathfrak{g}^{kn} \hat{j}_{mn}~,\]
allows us to identify the complex structure with a real selfdual form. The  latter are parameterized as follows, see appendix \ref{sec:torsion2},
\[ j_x = \alpha \hat{j}_x + \frac{\beta}{2} \hat{\omega}_x + \frac{{\beta}^*}{2} {\hat{\omega}}^*_x~,\]
where  $\alpha$ is real and $\alpha^2+|\beta|^2=1$. Hence the space of complex structures $I_x$ compatible with the metric $\mathfrak{g}_x$ forms a sphere 
whose coordinates  $\theta\in [0,\pi]$, $\psi\in [0,2\pi)$ are defined by $\alpha = \cos \theta$, $\beta= \sin\theta e^{i\psi}$, so that $I_x$ is 
associated with the two-form,
\[ j_x =\cos\theta\ \hat{j}_x + \sin\theta\ \Re(e^{i\psi}\hat{\omega}_x)~.\]
Extending this procedure to each point on $B_4$ then defines an almost complex structure $I$ over the whole manifold (unlike $\hat{I}$, 
$I$ will not be integrable in general). 
Over each point on $B_4$ an almost complex structure compatible with the metric of $B_4$ can be thought of as a point on the sphere $S^2$ parameterized by $(\theta,\psi)$.  
Hence the space of almost complex structures on $B_4$ is a fiber bundle $S^2$ over $B_4$ denoted by Tw$(B_4)$, the {\it twistor space} of $B_4$.

The  zweibein $z_1,z_2$ is no longer compatible with the almost complex structure $I$ associated with the real two-form $j$ given above. Rather we define,
\eq{\spl{\label{32}
f_1&:= \cos\frac{\theta}{2}e^{i\frac{\psi}{2}} z_1 + i\sin\frac{\theta}{2}e^{-i\frac{\psi}{2}}\bar{z}_2\\
f_2&:= \cos\frac{\theta}{2}e^{i\frac{\psi}{2}} z_2 - i\sin\frac{\theta}{2}e^{-i\frac{\psi}{2}}\bar{z}_1~,
}}
so that $I f_k=i f_k$. In terms of the new zweibein the local $SU(2)$ structure and the metric read,
\[\begin{array}{rcl}
\mathfrak{g}&=& f_1 \bar{f}_1+ f_2 \bar{f}_2    \\
j            &=& \frac{i}{2}\left( f_1 \wedge\bar{f}_1+ f_2\wedge \bar{f}_2\right) \\
\omega       &=& f_1\wedge f_2 = \cos \theta\ \Re(e^{i\psi}\hat{\omega}) -\sin\theta\ \hat{j} + i\ \Im(e^{i\psi}\hat{\omega})~,
\end{array}\]
which is precisely of the form of (\ref{rcl1}). 
Let us also note that the choice of zweibein 
compatible with $I$ 
is only determined up to a phase. The latter leaves $j$ and the metric invariant but acts nontrivially on $\omega$, thus changing the $SU(2)$ structure.

We have seen that  $I_x(\theta,\psi)$ defines an almost complex structure on the base. Together with
 the natural complex structure of the sphere (thought of as a $\cp{1}$) we can construct an almost complex structure on the the total space,
\[I_\pm=\left(\begin{array}{cc}
I_x(\theta,\psi) & 0_{4\times2} \\
0_{2\times4}     & \begin{array}{cc}                     0           & \pm\frac{1}{\sin\theta}\\        \mp\sin\theta & 0                \end{array}
\end{array}\right)~,\]
so that $f_1,f_2$ and $K = \d{}\theta +i \sin\theta (\d{}\psi +A)$ are eigenforms of $I_\pm$ with eigenvalue $\pm i$. We can thus take $(f_1,f_2,K)$ 
as the vielbein on Tw$(B_4)$. More generally we could modify $(f_1,f_2,K)$ by introducing ``warp factors'' as in (\ref{r2}) below.

\section{Torsion classes for K\"{a}lher base}\label{sec:apptorsionkaehler}

As mentioned in  section \ref{sec:lt} we may 
 relax the condition on the base of $S^2(B_4)$, so that $B_4$ is a generic four-dimensional K\"{a}hler manifold. The torsion classes can 
also be straightforwardly calculated in this case. Note however that this is only a local calculation: without additional constraints, we do not expect there 
to exist a global extension to a complete space. 

Let us postulate a globally-defined  $SU(3)$ structure as in (\ref{r2})  on a $\cp{1}$ bundle with metric, 
\begin{equation}
g_6= |h|^2 g_4 + KK^*~;~~~K=f \d{}\theta + ig(\d{}\psi+A)~,
\end{equation}
where $f,g,h$ are \textit{a priori} complex functions; $\theta$ and $\psi$ parameterize the $S^2$ fiber; the one-form $A$ satisfies (\ref{kaehlerclassesb}), (\ref{d5}) for 
$(J,\Omega)\rightarrow(\hat{j},\hat{\omega})$. 
We will impose further restrictions on $f,g,h$; these functions must be regular and non-vanishing, except for $g$ which must vanish at $\theta=0$ and $\theta=\pi$. The most general situation we will consider  here is that $\d{}f$, $\d{}g$, $\d{}h$ live on the space spanned by $K,K^*$ (this restricts the dependance on the coordinates). 
Explicitly we expand, 
\eq{\d{} f=f_1 K + f_2 K^*~,} 
and similarly for $g$, $h$. It is also possible restrict the dependance on $\theta$ alone.

The calculation of the torsion classes proceeds in the same fashion as in appendix \ref{sec:torsion}, with the following result,
\begin{equation}\label{oiu}
\begin{array}{rcl}
W_1 &=& -\frac{2i}{3}\frac{h}{h^*}\left(\frac{g+f\sin\theta}{fg^*+f^*g} +\frac{R}{2}g\frac{\sin\theta}{|h|^2}\right) \\
W_2 &=& \frac{2i}{3}\frac{h}{h^*}\left(\frac{g+f\sin\theta}{fg^*+f^*g} - R g\frac{\sin\theta}{|h|^2}\right)J^\perp\\
W_3 &=& -\frac{1}{2} (fg^*+f^*g)\d{}\theta\wedge\tilde{R}  + \Re \left(\frac{g-f\sin\theta}{fg^*+f^*g}  + \frac{R}{2} g\frac{\sin\theta}{|h|^2}\right)\Omega^\perp\\
W_4 &=& \d{} (\log |h^2|)-\frac{R}{2|h^2|}(fg^*+f^*g)\cos\theta\d{}\theta \\
W_5^* &=& \frac{1}{fg^*+f^*g} \left(f \cos\theta+f_1g-fg_1 -(f^*g_2+f_2g^*)- 2(fg^*+f^*g)\frac{h_2}{h}   \right) K^*~.
\end{array}
\end{equation}
Our degrees of freedom in the above are a somewhat redundant: a phase change of $K$ can be absorbed in $h$ so that $f$ or $g$ can be taken real. Let us also note that in general a cross term $(fg^*-f^*g)\d{}\theta (\d{}\psi+A)$ appears in the metric. If we want this to vanish, we must impose $f$ and $g$ to be colinear, so that they can both be taken real.

Furthermore  if we want to impose $W_4=0$, we must restrict $h$ to depend only on $\theta$, in which case we get,
\eq{
h_1 = \frac{g^* h'}{fg^*+f^*g}~;~~~ 
h_2 = \frac{g h'}{fg^*+f^*g}~.
}
Therefore $f$ and $g$ must also be restricted so that $R~\!(f^*g+fg^*)$ is a function of $\theta$ alone.

\vfill\break


%
%

\bibliography{refs}
\bibliographystyle{unsrt}
\end{document}